\documentclass[jcp,reprint,aip,superscriptaddress]{revtex4-2}
\usepackage[utf8]{inputenc}

\usepackage{graphicx}
\usepackage{amsmath}
\usepackage{xcolor}
\newcommand{\Angstrom}{\text{\AA}}
\newcommand{\sub}[1]{\ensuremath{_{\mathrm{#1}}}}
\newcommand{\super}[1]{\ensuremath{^{\mathrm{#1}}}}

\begin{document}
\title{Interfacial water asymmetry at ideal electrochemical interfaces}
\author{Abhishek Shandilya}
\affiliation{Department of Materials Science and Engineering, Rensselaer Polytechnic Institute}
\author{Kathleen Schwarz}
\email{kathleen.schwarz@nist.gov}
\affiliation{Material Measurement Laboratory, National Institute of Standards and Technology}
\author{Ravishankar Sundararaman}
\email{sundar@rpi.edu}
\affiliation{Department of Materials Science and Engineering, Rensselaer Polytechnic Institute}

\begin{abstract}
Controlling electrochemical reactivity requires a detailed understanding of the charging behavior and thermodynamics of the electrochemical interface.
Experiments can independently probe the overall charge response of the electrochemical double layer by capacitance measurements, and the thermodynamics of the inner layer with potential of maximum entropy (PME) measurements.
Relating these properties by computational modeling of the electrochemical interface has so far been challenging due to the low accuracy of classical molecular dynamics (MD) for capacitance and the limited time and length scales of \emph{ab initio} MD (AIMD).
Here, we combine large ensembles of long-time-scale classical MD simulations with charge response from electronic density functional theory (DFT) to predict the potential-dependent capacitance of a family of ideal aqueous electrochemical interfaces with different peak capacitances.
We show that, while the potential of maximum capacitance varies, this entire family exhibits an electrode charge of maximum capacitance (CMC) between -3.7 $\mu$C/cm$^2$ and -3.3 $\mu$C/cm$^2$, regardless of details in the electronic response.
Simulated heating of the same interfaces reveals that the entropy peaks at a charge of maximum entropy (CME) of $-6.4 \pm 0.7~\mu$C/cm$^2$, in agreement with experimental findings for metallic electrodes.
The CME and CMC both indicate asymmetric response of interfacial water that is stronger for negatively charged electrodes, while the difference between CME and CMC illustrates the richness in behavior of even the ideal electrochemical interface.
\end{abstract}

\maketitle

\section{Introduction}
Chemistry at the electrochemical interface underpins a wide range of energy conversion,\cite{RevEnergyConversion} energy storage\cite{RevEnergyStorage,Goodenough} and chemical synthesis technologies.\cite{RevChemSyn}
An important feature of electrochemical processes is their sensitivity to electrode potential, which provides an additional mode of control not available for reactions in the liquid phase and at solid-gas interfaces.
Designing electrodes and electrolytes to fully exploit this control and target specific chemical reactions requires a comprehensive understanding of the thermodynamics and atomic-scale charge response of the electrochemical interface.

Measurements of electrochemical capacitance as a function of electrode potential provide a sensitive experimental probe of the overall charge response of the electrochemical double layer.
However, the capacitance depends on the solvent dielectric response in the inner layer, ionic response in the diffuse layer, and specific adsorption of ions.
It is not straightforward to disentangle these effects, especially near the potential of zero charge (PZC).
The capacitance near the PZC is dominated by the low capacitance of the diffuse layer at the low ionic concentrations typically used in experiments.
The capacitance at potentials far from the PZC is determined by a combination of dielectric saturation in the inner layer and ion packing effects.
It remains unclear from experiments whether the inner layer capacitance (after removing the diffuse layer capacitance dip) exhibits a potential of maximum capacitance (PMC) coincident with the PZC, or away from the PZC, indicating an asymmetric charge response of the inner layer.
This inner layer response is particularly important for the energetics of chemical reactions at the surface, and it is therefore critical to disentangle it from the diffuse layer and ion effects.

A complementary probe of electrochemical interfaces that is more sensitive to the inner layer response is the potential of maximum entropy (PME),\cite{Pt111PME2} measured from the temperature dependence of the electrode-electrolyte potential difference, $V$, at fixed charge density.\cite{HgPME}
Specifically, the electrode charge density $\sigma$ at which the potential does not change with temperature corresponds to the charge of maximum entropy (CME), because the thermodynamic relation $\partial V/\partial T|_\sigma = -\partial S/\partial \sigma|_T$ implies that $\partial S/\partial \sigma = 0$ at this point.
(The corresponding potential is the PME.)
This is most cleanly realized by measuring potential transients following laser-induced heating of the interface, as this minimizes other effects of a temperature change.\cite{laserPME, PMEElectrolytes, KoperNiOH, Pt111PME2, PtBiPME}
The PME is typically close to and slightly below the PZC for metal electrodes in aqueous electrolytes,\cite{HgPME, laserPME, FeliuPME, FeliuPMEPt}, e.g., $0.1$~V below PZC for Ir(111),\cite{IrPME} allowing it to be used as an approximate measure of the PZC.\cite{CuPME}
The corresponding charge (CME) is approximately $-5~\mu$C/cm$^2$ for Au(111),\cite{laserPME} and  between ($-4$ to $-6$) $~\mu$C/cm$^2$ for mercury.\cite{HgPME}
The apparently universal negative CME in metal-water interfaces is attributed to the oxygen end of water facing the electrode at the PZC, requiring a negative electrode charge to counter this preferential orientation and increase the entropy.\cite{laserPME} 
This asymmetry of interfacial water also agrees with recent spectroscopic evidence\cite{AsymmWater} and molecular dynamics (MD) studies on interfacial water dynamics~\cite{Dynamics}, prompting the question whether the inner layer capacitance is similarly asymmetric.

In particular, do the charges of maximum entropy (CME) and capacitance (CMC) coincide?
These are thermodynamically distinct quantities, with CME corresponding to $\partial V/\partial T|_\sigma = 0$ and the CMC to $\partial^2 V/\partial\sigma^2|_T = 0$, and could be different in general.
In the simplest formulations of an asymmetric continuum dielectric, these quantities may be expected to coincide.
For example, if the nonlinear dielectric response $\epsilon(\mathcal{E})$ is assumed to peak at a non-zero electric field $\mathcal{E}$ at the interface, instead of at zero field like the bulk response, then the dipole-orientation entropy will also peak at the same field.
This is because the nonlinearity of the dielectric response stems primarily from the competition between the potential energy of the molecular dipoles in the electric field and the entropy of dipole reorientation.\cite{PolarizableCDFT}
In this case, the capacitance and entropy of the solvent layer at the interface will peak at the same interfacial field $\mathcal{E}$, and hence at the same surface charge density $\sigma$.
However, this simplified picture does not account for capacitance and entropy contributions from adsorption or electron transfer effects, or from beyond the solvent layer (e.g., ions), which could lead to differences between the CMC and CME.
Consequently, evaluating the relationship between CMC and CME will be invaluable in developing simplified models of the electrochemical interface, such as continuum solvation models for first-principles electrochemistry,\cite{CANDLE} that correctly account for the asymmetry in the charge response of the interface.

Computational prediction of capacitance and entropy of electrochemical interfaces under identical conditions would provide great insight into the relation between the charge response and thermodynamics of the interface.
However, this has been challenging due to limitations of MD simulations that can address both these properties.
\emph{Ab initio} MD (AIMD) simulations can capture all relevant physical effects by density-functional theory (DFT) treatment of the electrons in principle, but are limited in time and length scales required to accurately model the capacitance of the interface.
Classical MD simulations can achieve the required scales to model electrochemical interfaces,\cite{NPTEchem, GrapheneAqueousElectrolyte, IonicLiquidDLformation, ChargeFluctIons, ChargeFluctElectrons, Chandler} but require special care for capacitance predictions.\cite{CapacitanceIssues1, CapacitanceIssues2}
In particular, such simulations must account for the electronic response of the electrode extending past the surface atoms,\cite{Bokris} which can be done by shifting the `effective electron-response plane' based on separate DFT calculations or by incorporating simplified electron-response models such as Thomas-Fermi screening.\cite{PotDrop, ThomasFermiMetallic}
Such techniques have been applied to electrochemical capacitance with ionic liquid electrolytes,\cite{KornyshevElectronPlane, RuzanovIonicLiquidCap, PaekCombineDFT-MD, PotDrop} but infrequently for metallic electrodes with aqueous electrolyte.\cite{adamPot}
Predicting the entropy of the electrochemical interface to evaluate PME or CME has remained even more challenging.
Pioneering attempts based on analyzing fluctuations of the work function in \emph{ab initio} MD simulations\cite{Rossmeisl} have been limited in their quantitative comparison to experimental PME due to computational cost and the difficulty in referencing the electrochemical potential.\cite{Cheng}

In this Article, we combine classical MD simulations of ideal aqueous metal-electrode interfaces with both the effective electron-response plane approach,\cite{KornyshevElectronPlane, PotDrop} and the electronic response from DFT calculations.
Prediction of absolute capacitance for a specific aqueous metal electrode interface still remains a challenge, so we instead predict a family of potential-dependent capacitance curves corresponding to interfaces with different peak values of capacitance.
We find that the potential of maximum capacitance (PMC) relative to PZC is always negative and depends on the peak capacitance value, but that the charge of maximum capacitance (CMC) is constant across the family and depends only slightly on the technique used for incorporating the electronic response.
We then compare the CMC to the charge of maximum entropy (CME), predicted by directly simulating the heating of the same electrochemical interfaces in MD.
We show that the CMC and CME are both negative, indicating asymmetric response of water with the same sign for charge response and thermodynamics, but that their magnitudes are distinct at a level well above the accuracy of our predictions.

\section{Methods}

The capacitance and entropy of real electrochemical interfaces can be strongly affected by several effects that depend on specific electrodes and electrolytes.
In particular, adsorption at the interface can occur to a varying degree for electrolyte ions and water molecules.
Experimentally, it can be hard to eliminate ion adsorption pseudocapacitance contributions to the measured capacitance in situations of high surface coverage of adsorbates.\cite{Valette} Additionally, strongly hydrophilic surfaces may adsorb water and alter the surface dipole.\cite{Cheng, KoperPt}
Here we target an ideal electrochemical interface without any such effects that add complexity to the interpretation of the CMC and CME.

\subsection{Capacitance calculation overview}
\label{sec:CapacitanceMethod}

\begin{figure}
\includegraphics[width=\columnwidth]{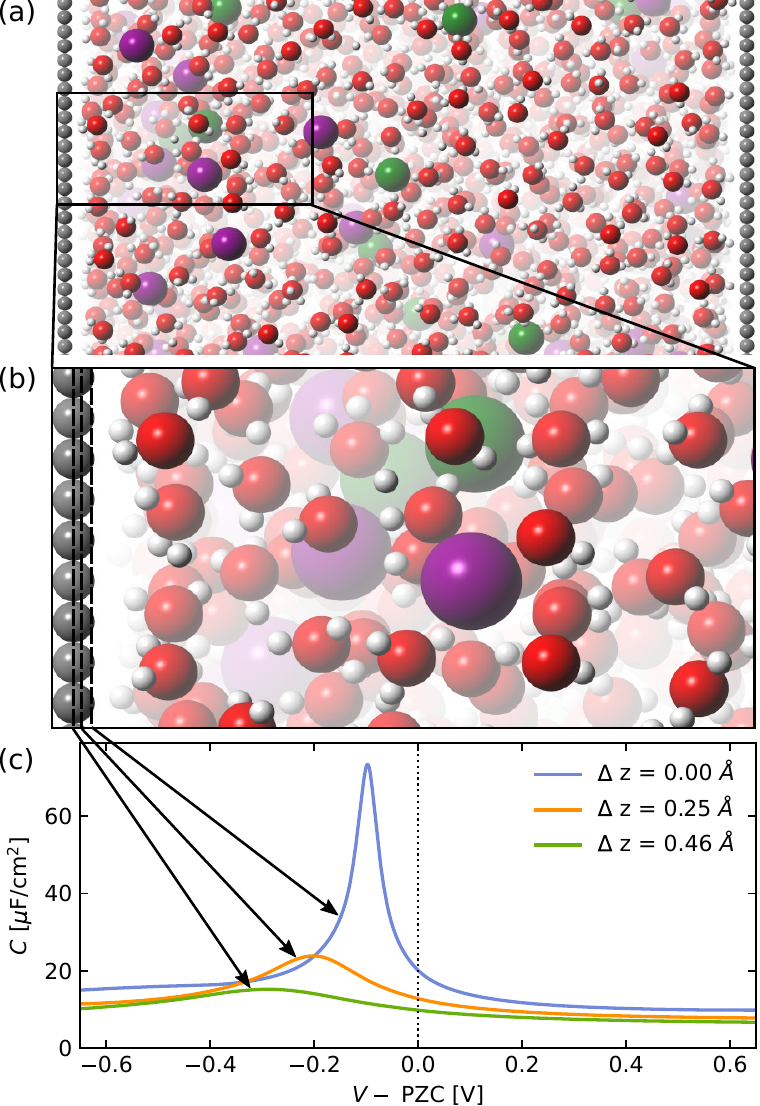}
\caption{(a) Typical snapshot of the classical molecular dynamics (MD) simulation containing two back-to-back half-cells with equal electrode charge, and (b) a closer view of one half cell of Ag(100) in 1~mol/L aqueous NaF.
(c) The capacitance predicted directly from the unmodified MD electrode charge density at the plane of atoms is too low, requiring a shift of the effective electrode response plane towards the electrolyte to account for electronic response absent in classical MD.
We set $\Delta z = 0$ such that the capacitance at PZC is $20~\mu$F/cm$^2$, and analyze the behavior of a family of ideal metal-water interfaces with varying peak capacitances by changing $\Delta z$.}
\label{fig:schematic}
\end{figure}

FIG.~\ref{fig:schematic}(a) and (b) show a typical snapshot of our overall classical MD simulation cell (details in Section~\ref{sec:DetailsMD} below) and a close-up of the interfacial region.
With charges on the surface metal atoms and the closest H atoms in water separated by $d > 1$~\AA, these configurations contain a  `vacuum' capacitance contribution in series of $\epsilon_0/d \approx 8.8~\mu$F/cm$^2$, where $\epsilon_0$ is the permittivity of vacuum.
The direct capacitance prediction from classical MD must therefore be smaller than this value, as indeed seen in the lowest curve in FIG.~\ref{fig:schematic}(c).
This is much smaller than the typical approximately $ 50~\mu$F/cm$^2$ peak double-layer capacitance of aqueous metal electrodes.\cite{KoperPt,Valette}

In a real electrochemical interface, the electronic response of both the metal and water extend significantly past the planes of the corresponding atoms, substantially narrowing the effective vacuum gap $d$.
This effect can be captured in principle by \emph{ab initio} MD simulations with full quantum-mechanical treatment of all electrons, but the nanosecond time scale of ion equilibration is computationally prohibitive, especially for unit cells with thousands of atoms that include a statistically significant number of ions.
DFT calculations of metal-water interfaces with frozen water geometries are feasible to calculate electron spill-over effects, but miss the dielectric response from solvent-dipole reorientation.
Such calculations are therefore difficult to combine with other models to predict the overall interfacial capacitance.

Consequently, we adopt the effective electron-response plane approach\cite{KornyshevElectronPlane} to narrow the vacuum gap $d$ by virtually moving the electrode charge location towards the electrolyte.
As long as the electrode and electrolyte charge densities do not overlap, this does not change the electric field due to the electrode in the electrolyte region by Gauss's law.
Therefore, this change does not alter the MD trajectories and amounts to measuring the electrostatic potential from the MD simulation at a modified location in the final analysis.
In principle, DFT can predict the effective charge density response location, and this works reasonably for graphene-water interfaces.\cite{PotDrop}
However, for metal-water interfaces with significantly higher capacitance, the effective gap is nearly zero, leading to extreme sensitivity of the predicted capacitance to the charge location.

To circumvent this issue, we focus not on the absolute capacitance of a specific metal-water interface, but the behavior of the potential-dependent capacitance for a family of ideal metal-water interfaces with different charge response locations.
Further, the location of the metal atoms in the MD simulation is no longer relevant in the analysis of the capacitance: they serve primarily to set up an interfacial potential and electric field for the electrolyte.
Similarly, there is no particularly meaningful or stable spatial location in the liquid profile to reference the plane location.
Therefore, we reference the electron response plane location based on the predicted capacitance curves, allowing us to more conveniently compare the trends between different methods below.
Specifically, we pick the reference $\Delta z = 0$ for the effective electron response plane to be the location that yields a capacitance $C\sub{PZC} = 20~\mu$F/cm$^2$ at the PZC.
We pick this value because it leads to overall capacitance values typical for metal-water interfaces and does not lead to appreciable overlap with the electrolyte charge density.
With $\Delta z = 0$ as the maximum capacitance curve, we calculate the capacitance for several values of $\Delta z > 0$ moving the electrode charge away from the electrolyte.
Note that at just $\Delta z = 0.25$~\AA, the peak capacitance already drops to $25~\mu$F/cm$^2$ (Fig.~\ref{fig:schematic}(c)), smaller than for most metal-water interfaces, while the unmodified MD charge density corresponds to a $\Delta z \approx 0.46$~\AA.

In summary, we predict a family of capacitance curves for ideal metal-water interfaces by varying the location of the effective electrode charge response location indexed by $\Delta z$ based on a specific value of capacitance at PZC.
Note that the water molecules are free to move closer to the electrode with increasing electric field magnitude (in both directions away from PZC), therefore accounting for any electrostriction effects that are particularly important for highly  compressible fluids such as ionic liquids,\cite{Estrict} and that have also been observed in confined aqueous electrolytes.\cite{Electrostriction}
We also investigate the effect of the charge-dependent metal electronic response by directly combining electron density profiles from DFT with the MD charge density (Section~\ref{sec:DetailsDFT}).

\subsection{MD simulation details}\label{sec:DetailsMD}

We use the Large-scale Atomic/Molecular Massively Parallel Simulator (LAMMPS\cite{LAMMPS}) to perform classical MD simulations of aqueous 1 mol/L NaF electrolyte between Ag(100) electrodes in a $45\times 45\times 60~\Angstrom^3$ unit cell with periodic boundary conditions.
The chosen ionic concentration is deliberately high compared to typical electrochemical experiments for two reasons: (1) the diffuse layer capacitance is less significant at this concentration allowing us to focus on inner layer properties, and (2) lower concentrations will require larger unit cells and present greater statistical sampling challenges.
We pick NaF as the simplest of nominally non-adsorbing electrolytes, although F$^-$ may exhibit greater specific adsorption than more complex compound ions such as ClO$_4^-$ or KPF$_6^-$.
The space between the electrodes is $45\times 45\times 43.6~\Angstrom^3$, of which FIG.~\ref{fig:schematic}(a) shows a section with approximately 10~\Angstrom~depth visible into the plane of the page.

Each simulation includes two back-to-back half cells, with more than enough distance to separate the two diffuse layers, given that the screening length is $\sim 3$~\AA~for this electrolyte at 298~K (using the Debye screening length as a rough estimate).
We set up the same charge in both half cells to create nominally inversion-symmetric unit cells, avoiding issues with long-range dipole interactions and ion equilibration between the two cells.
The electrodes are treated with a single layer of charged atoms with effective potentials capturing the interaction of the electrolyte with an Ag(100) slab as discussed below, and are separated by 14~\Angstrom~vacuum, found to be sufficient to suppress the interaction between periodic images of the electrolyte.
For each randomly-initialized configuration discussed below, we perform energy minimization followed by NVT simulations at 298~K with a 2~fs time step, discard the first 1~ns for equilibration, and capture statistics over 9~ns.

We use the rigid extended simple point charge model (SPC/E) water model\cite{SPCE} with molecule geometry constrained by the the SHAKE algorithm,\cite{SHAKE1, SHAKE2} and with Lorentz-Berthelot mixing of Lennard-Jones parameters (arithmetic for $\sigma$, geometric for $\epsilon$) except between the Na$^+$ and F$^-$ ions for accurate treatment of the ion-pair interactions.\cite{Fyta2012}
The SPC/E water model underestimates the dielectric constant of water and the air/water surface tension by 10 -- 20\%,\cite{PolarizableCDFT} but captures their trends with temperature and should suffice for our exploration of asymmetric charge response.\cite{SPCESurfaceTension}
We parameterize a Morse potential for the short-ranged interaction between the electrode atoms and water / ions from electronic DFT calculations of a single molecule / atom next to a neutral Ag(100) surface, as detailed in the Supplementary Information.
This effective interaction parameterized to DFT includes the image-charge attraction between water molecules and the wall.
We therefore do not explicitly require a fixed-potential treatment of the metal electrode to capture this effect,\cite{Salanne} but neglect dependence of the electrode-electrolyte interaction on potential by using fixed metal-atom charges.
Note that while we require specific metal atom parameters for the MD simulation, we analyze the results more generally for the properties of ideal metal-water interfaces as discussed previously.

We initialize the simulation cell with a random distribution of water molecules created with using a simple Monte Carlo insertion method that enforces a minimum O-O separation of 2~\Angstrom.
We then randomly replace a selected number of water molecules with Na$^+$ or F$^-$ ions, one each from equally sized bins in the $z$ direction; this ensures a uniform initial spatial distribution of ions to mitigate ion equilibration times.
Finally, we picked the number of water molecules and ions iteratively based on trial MD simulations to ensure a bulk density of 1~g/cm$^3$ for water and 1~mol/L for NaF far in the center of the simulation cell.
For the neutral electrode we end up with 2809 water molecules and 46 Na$^+$ and F$^-$ ions each.

We charge the electrodes in steps of one additional ion of the same type (either Na$^+$ or F$^-$) in each half cell, amounting to a step of $1~e^-/(45~\Angstrom)^2 \approx 0.79~\mu$C/cm$^2$, with a compensating charge distributed equally among all the surface metal atoms.
We extend these simulations up to 20 extra ions of each type, thereby spanning electrode charges from $\sigma \approx (-16$ to $+16$) $~\mu$C/cm$^2$.
While the difference in ion numbers is fixed by charge neutrality of the simulation cell, the total number of ions may vary with the electrode charge.
Grand canonical simulations could ensure that the density of ions approaches the target bulk value (1~mol/liter each) in the center of the simulation cell, but are challenging to perform at the scale required here.
We find that fixing the number of minority ions with the same charge as the electrode (at 46 in our case), and increasing the number of majority ions of opposite charge as the electrode (as $46 + 2|\sigma|/$(0.79~$\mu$C/cm$^2$), keeps the bulk ionic concentration close to the targeted value.

For each of the 41 electrode charge values (including neutral), we perform five independent MD simulations starting from different random configurations, yielding 10 half cells for each charge point.
We then compute the planarly-averaged charge density profile (as a function of $z$) with the electrode charge density offset by different amounts as discussed in Section~\ref{sec:CapacitanceMethod}.
For each offset, we solve a 1D Poisson equation to get the electrostatic potential profile for each electrode charge, measure the potential $V$ between the electrode and the bulk region of the electrolyte.
Note that this is exactly equivalent to planarly-averaging the 3D Poisson equation because the Coulomb kernel and planar average are both diagonal operators in reciprocal space, and hence commute with each other.
For capacitance calculations, we only need differences in the potential far from the interface and therefore do not need to worry about short-ranged contributions to the local potential seen by an ion near the interface (Madelung potential).\cite{ReedMadelung}
Additionally, the interface potential difference, $V$, that we calculate from a point charge model will differ by an overall constant compared to more realistic charge distributions of the water molecules and ions,\cite{WaterSurfPot} but this does not impact the differences in potential used in $V$ - PZC and the differential capacitance evaluation.
Finally, we compute $C = \partial\sigma/\partial V$ from a cubic spline fit of the $(\sigma, V)$ data to obtain a differential capacitance curve. (See FIG.~\ref{fig:capExtraction} in Appendix~\ref{sec:AppendixMethods} for details.)
Note that we perform each charge simulation from completely independent random configurations, and therefore the smoothness of the obtained charging curves confirms adequate equilibration of our simulations.

Finally, to evaluate the charge and potential of maximum entropy, we repeat the entire set of classical MD simulations above at 318~K and compute $dV/dT$ for each electrode charge density $\sigma$ from finite-difference derivatives between (298 and 318)~K.
Further details on the analysis of these results are discussed below in Section~\ref{sec:ResultsEntropy}.

\subsection{Electronic DFT details}\label{sec:DetailsDFT}

So far we discussed capacitance prediction from the classical MD simulations by offsetting the electrode charge location to account for electronic response.
We also present results below that include the electronic response of the electrode from DFT.
We perform electronic DFT calculations of a 7-layer Ag(100) surface in the JDFTx code,\cite{JDFTx} with the Perdew-Burke-Ernzerhof (PBE) exchange-correlation functional,\cite{PBE} ultrasoft pseudopotentials\cite{GBRV} at kinetic energy cutoffs of 20 Hartrees for the wavefunction and 100 Hartrees for the charge density, a $12\times 12 \times 1$ $k$-point mesh and Fermi smearing of 0.01~Hartrees.
The slabs are separated by 16~\AA~vacuum, and truncated coulomb potentials are used to remove periodic interaction between the slabs.\cite{TruncatedEXX}
We apply electric fields perpendicular to the surface from 0 to 17.8~V/nm in steps of 0.89~V/nm (21 calculations) to obtain equal and opposite surface charge densities in steps of $0.79~\mu$C/cm$^2$ matching the MD simulations above.

We extract the difference in electron density from the zero-field simulation as the DFT electrode charge density profile for each of the 41 surface charge densities simulated in MD (zero and 20 charge magnitudes of each sign).
The above DFT simulations implicitly include the nonlinear response of the electrode to surface charge density.
To test the effect of the nonlinearity, we also calculate the linear-response change in electron density to an infinitesimal electric field using density-functional perturbation theory (DFPT), with all other parameters identical to the DFT calculations above.
We then replace the planarly-averaged charge density contribution from the classical MD electrode with these electronic charge density profiles (both from DFT and DFPT) before solving Poisson equation for the potential to analyze the impact of electronic charge response on the capacitance.

\begin{figure*}
\includegraphics[width=\textwidth]{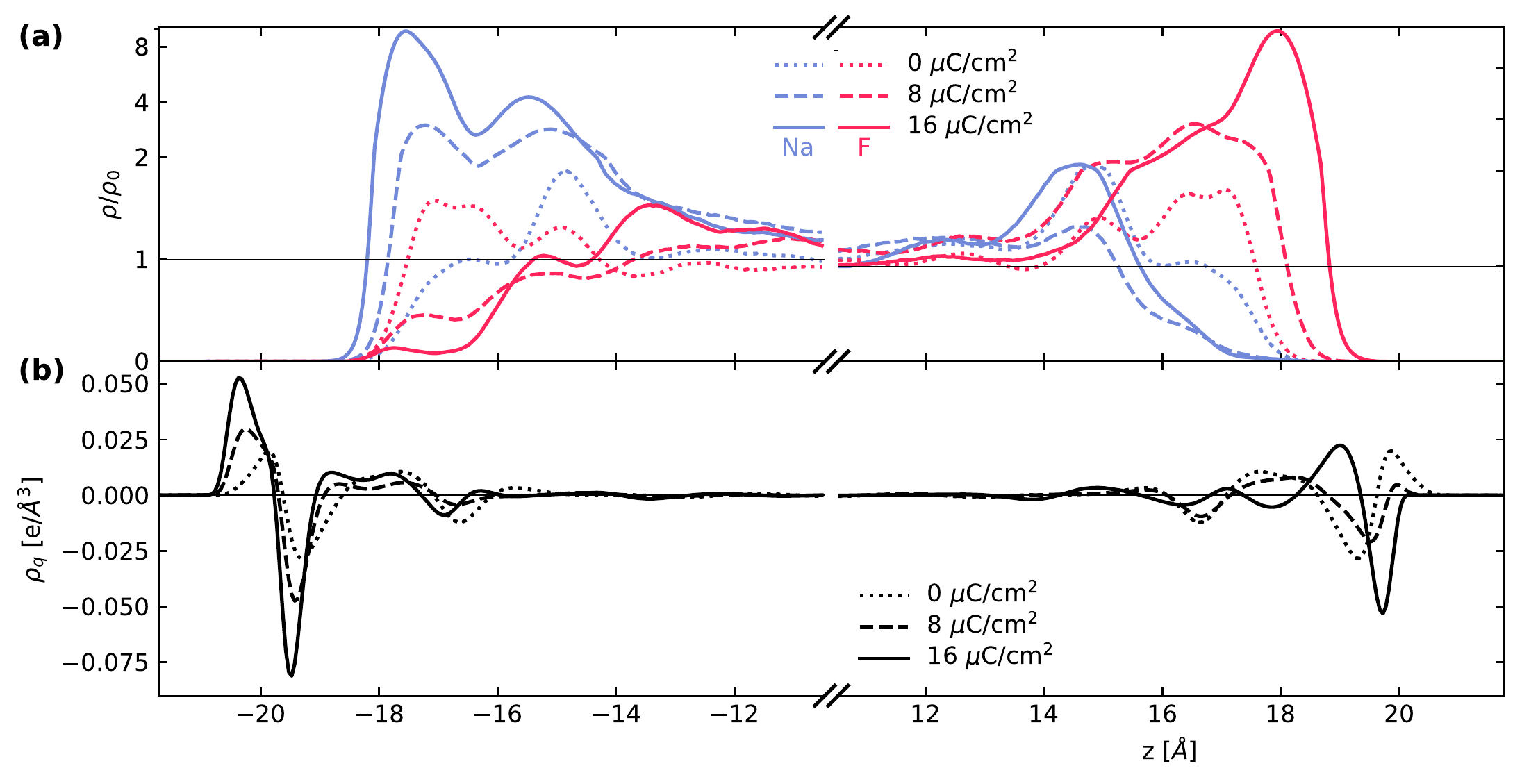}
\caption{(a) Ion density profiles and (b) total electrolyte charge density for different electrode charge densities.
The break in the x-axis separates results for negatively-charged electrodes on the left and positively-charged electrodes on the right, averaged over 10 half cells for each charge.
The charge density closest to the electrodes is entirely from water, and is larger in magnitude and nearer to the negative electrodes, leading to a higher capacitance for negative potentials.}
\label{fig:charge-ions}
\end{figure*}

\section{Results and Discussion}

\subsection{Charge density profiles} \label{sec:ResultsCharge}

We begin with an analysis of the variation of electrolyte charge distributions with electrode charge, as predicted from ensembles of long-time classical MD simulations.
FIG.~\ref{fig:charge-ions} compares ion density and total charge density profiles averaged over 10 half cells each for various electrode charges, with negative electrode charges on the left and positive electrode charges on the right.
Note that the results on the left and right are from separate simulations: each MD simulation contains electrodes of the same sign to avoid overall unit cell dipoles, as discussed in Section~\ref{sec:DetailsMD}.

Each ion density in FIG.~\ref{fig:charge-ions}(a) increases substantially from the bulk values in the vicinity of the oppositely charged electrode -- Na$^+$ near the negative electrodes on the left and F$^-$ near the positive electrode on the right, as expected.
Correspondingly, the ion densities are suppressed near the like-charged electrode.
The peak ion densities are approximately 4~\Angstrom~away from the electrode surfaces (at $\pm21.8$~\Angstrom~in $z$).
In another (2 -- 3)~\Angstrom~further, the ion profiles transition to a rapid decay towards the bulk density, as expected.

Importantly, the ion profiles are not equal for the neutral electrode: there is a small excess of F$^-$ closer to the electrode, with a small peak of Na$^+$ further out.
We would intuitively expect such profiles for a slightly positively-charged electrode, but instead find it for a neutral electrode.
Consequently, the ion profiles we expect for a neutral electrode would instead appear for slightly negatively-charged electrodes.
This gives a first indication that properties we expect to be symmetric about zero charge from classical continuum models, such as dielectric response and capacitance, might be centered at a negative charge instead.

The ion densities discussed above are further from the electrode than the first layer of water, and the net charge density profiles of the electrolyte shown in FIG.~\ref{fig:charge-ions}(b) is dominated by water for the first 3~\Angstrom~from the electrode.
In particular, the charge density adjacent to a neutral electrode starts with a nearer positive H peak, followed by a negative O peak further away, in agreement with AIMD predictions for metallic surfaces.\cite{ChengReview}
When the electrode is charged positively, the nearer H peak is suppressed in magnitude and the further-away O peak is enhanced.
In contrast, for negative electrode charges, the strength of both the H and O peaks is enhanced, leading to a larger charge density response than the positive case.
Most importantly, the charge response is closer to the electrode on the negative side: this should lead to a smaller potential difference for the same charge magnitude, and hence a larger capacitance on the negative side.

\subsection{Capacitance} \label{sec:ResultsCapacitance}

\begin{figure}[htp!]
\includegraphics[width=0.9\columnwidth]{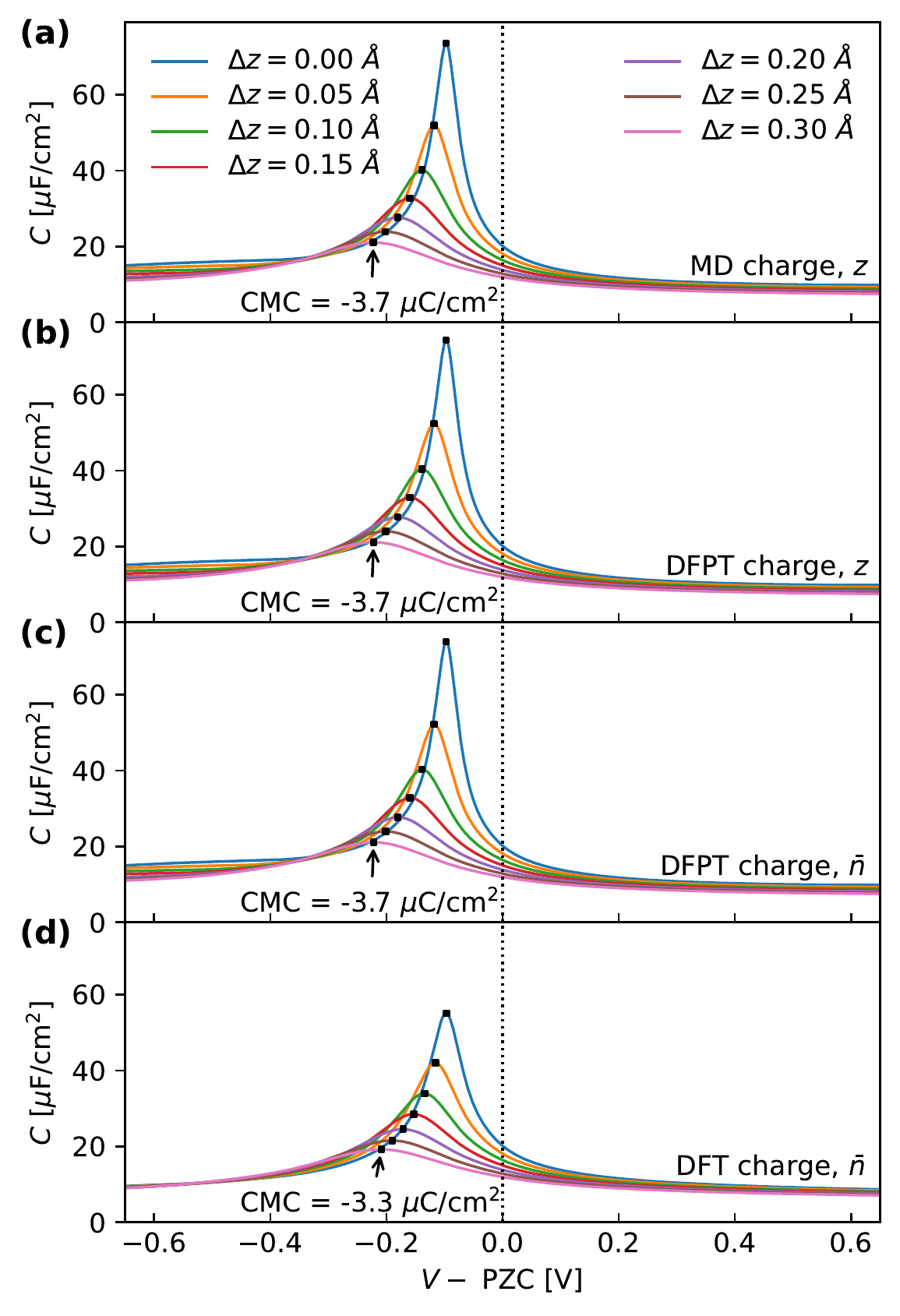}
\caption{Family of ideal aqueous electrochemical interface calculates using different models of the electrode charge density:
(a) classical MD point charges, (b) DFPT (linear response of DFT) charge density placed at a specific $z$, (c) DFPT charge density placed based on electrode-water electron density overlap, $\bar{n}$, and (d) DFT charge density (nonlinear response) placed by $\bar{n}$.
The family of capacitance curves is generated by offsetting the electrode charge model by $\Delta z$, with $\Delta z = 0$ set for each such that $C\sub{PZC} = 20~\mu$C/cm$^2$.
(FIG.~\ref{fig:rho} shows the corresponding electrode and electrolyte charge distributions for $\Delta z = 0$.)
The potential of maximum capacitance (PMC) varies from -0.09 to -0.22~V from the PZC across the family, but the charge of maximum capacitance (CMC) is constant at $-3.7~\mu$C/cm$^2$ in (a-c) and $-3.3~\mu$C/cm$^2$ in (d).}
\label{fig:capacitance}
\end{figure}

\begin{figure}[htp!]
\includegraphics[width=\columnwidth]{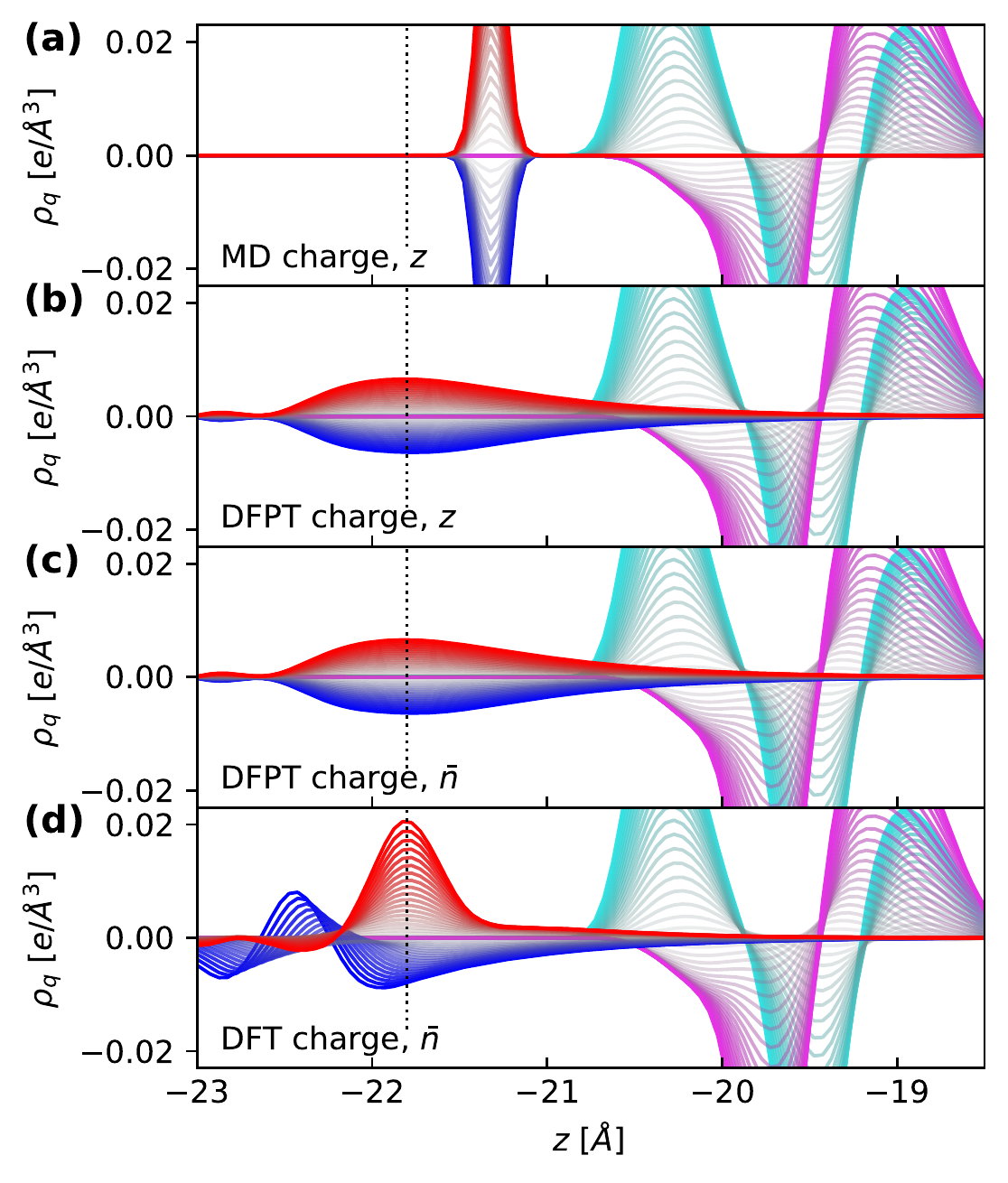}
\caption{
Electrode charge densities (41 curves on the left from blue for $\sigma \approx -16~\mu$C/cm$^2$ to red for $+16~\mu$C/cm$^2$) for each of the charge models and placement schemes at $\Delta z = 0$ in FIG.~\ref{fig:capacitance}, and corresponding electrolyte charge densities (41 curves from cyan to magenta on the right).
The vertical dotted line marks the location of the classical-MD metal atoms.
In (a), the electrode charge is a $\delta$-function in the calculation, broadened to a Gaussian with width 0.01~\AA~only for representation on the plot and strictly does not overlap with the electrolyte charge.
Charge density overlap is non-zero but small in (b-d), resulting in negligible changes to the capacitance in FIG.~\ref{fig:capacitance}(b,c) compared to FIG.~\ref{fig:capacitance}(a).
Only the asymmetry of the nonlinear electron density response between negative and positive charges in (d) visibly modifies the capacitance curve in FIG.~\ref{fig:capacitance}(d).}
\label{fig:rho}
\end{figure}

As discussed in Section~\ref{sec:ResultsCharge}, we expect the capacitance of the interface to peak at negative electrode charges, consistent with the predicted capacitance curves shown in FIG.~\ref{fig:schematic}(c).
Next, we turn to a quantitative analysis of the capacitance asymmetry and the location of its maximum.
FIG.~\ref{fig:capacitance} shows the family of capacitance curves for ideal aqueous electrochemical interfaces with different peak capacitances obtained with different models of the electrode charge density and different schemes of positioning the electrode charge density relative to the electrolyte charge density from MD, with the corresponding charge densities at the reference position, $\Delta z$, shown in FIG.~\ref{fig:rho}.

First, FIG.~\ref{fig:capacitance}(a) shows the direct prediction from MD after offsetting the electrode charge density location by different $\Delta z$, as detailed in Section~\ref{sec:CapacitanceMethod}.
The potential of maximum capacitance (PMC) is always negative, but its magnitude is inversely proportional to the peak capacitance value (see FIG.~\ref{fig:capPeak} in Appendix~\ref{sec:AppendixResults}).
Instead, if we look at the electrode charge density at the PMC, it is constant across the family, resulting in a charge of maximum capacitance (CMC) of $-3.7~\mu$C/cm$^2$ (with an uncertainty $\sim 0.2~\mu$C/cm$^2$ as estimated in FIG.~\ref{fig:capExtraction} in Appendix~\ref{sec:AppendixMethods}).

We can understand this behavior by noting that the asymmetry in the response of the water is a built-in polarization for the neutral electrode.
For a specific value of interfacial electric field $E$, this built-in polarization is neutralized, and we can expect this point to be the location of maximum capacitance.
This interfacial electric field is $E = \sigma/\epsilon_0$ by Gauss's law, directly determined by Gauss's law, while the electrode potential depends on the overall electrostatic potential profile and the electrode charge response location ($\Delta z$ in particular).
Hence, the CMC should be invariant across the family of capacitance curves, while the PMC depends sensitively on the overall capacitance.
Finally, note that the capacitance increases slightly for large potentials: this is due to an increased density of electrolyte in response to high electric fields at the interface (electrostriction).\cite{PolarizableCDFT}
The magnitude of this effect will be sensitive to the electrode-electrolyte interaction potential, and may be obscured by ion adsorption at potentials far from the PZC in experiment anyway.
Consequently, we focus on the asymmetric behavior of the capacitance close to the PZC below.

We next consider the impact of deviations from the simple point charge model of electrode charge density (sheet charge in the planar average) considered so far.
First, we use the linear-response charge density profile from DFPT (see Section~\ref{sec:DetailsDFT}) with a constant shape scaled to each electrode charge density value (FIG.~\ref{fig:rho}(b)).
FIG.~\ref{fig:capacitance}(b) shows the result of using this charge density profile instead of the MD charge, but with the same placement scheme as above: at specific locations in $z$ (with various $\Delta z$ offsets).
We find absolutely no difference in the capacitance curves and CMC, which can be explained by Gauss's law as long as the electrode and electrolyte charge densities do not overlap.
The exponential tail of the electronic charge response of the electrode does in fact overlap partially with the electrolyte charge density for the highest-capacitance $\Delta z = 0$ case (FIG~\ref{fig:rho}(b)), but this impacts the capacitance negligibly.
The overall potential difference for a given electrode charge density shape would match that for the sheet charge case when the sheet charge is placed at the center-of-charge of the charge distribution.
This center-of-charge location is absorbed into our definition of $\Delta z$ based on the $C\sub{PZC} = 20~\mu$C/cm$^2$ criterion (Section~\ref{sec:CapacitanceMethod}), and so the family of capacitance curves remains unchanged.

Next, let's account for the actual variation of electron density of the electrode.
First, consider the effect of the electron density on the short-ranged potential of the liquid: an increased electron density would lead to higher repulsion that pushes non-bonded liquid atoms away.
FIG.~\ref{fig:capacitance}(c) includes this effect by placing the electrode and electrolyte charges based on the overlap $\bar{n}(\vec{r}) = \int d\vec{r}' n(\vec{r}') n\sub{water}(\vec{r} - \vec{r}')$ of a water molecule's electron density $n\sub{water}(\vec{r})$ and the electrode electron density $n(\vec{r})$, as parameterized in the SaLSA solvation model.\cite{SaLSA, CANDLE}
Specifically, the separation at which $\bar{n}$ crosses $n_c = 1.42\times 10^{-3} a_0^{-3}$ correlates with the non-bonded distance of nearest approach.\cite{SaLSA}
We place the DFPT electrode charge density relative to the MD electrolyte density profiles based on this condition \emph{for each electrode charge}, and the offset by various $\Delta z$ to obtain the family of capacitance curves.
Interestingly, we find that the charge density profiles in FIG.~\ref{fig:rho}(c) are unchanged from the previous case.
Correspondingly, the capacitance curves and CMC still do not change (FIG.~\ref{fig:capacitance}(c)), indicating a negligible effect of the change in short-ranged repulsion with electrode charge density.

Finally, we use the full nonlinear variation of electrode charge density profile from DFT in FIG.~\ref{fig:capacitance}(d), which leads to a small but noticeable difference in the capacitance curves and CMC.
In particular, the magnitude of the CMC and the asymmetry overall is reduced compared to all previous cases.
Essentially, in the nonlinear response of the DFT, electron repulsion makes it is easier to positively charge the electrode by removing electrons than to negatively charge it by adding electrons.
This effect favors higher response on the positively-charged side (FIG.~\ref{fig:rho}(d)), and therefore reduces the overall asymmetry towards negative charges due to the water at the interface.
The net result is still a negative CMC, but reduced slightly in magnitude to $-3.3~\mu$C/cm$^2$.

\subsection{Entropy} \label{sec:ResultsEntropy}

Above, we showed that the capacitance peak occurs for negatively-charged electrodes with a CMC of $-3.7~\mu$C/cm$^2$ based on the MD charge densities, which reduces in magnitude to $-3.3~\mu$C/cm$^2$ accounting for nonlinearities in the electronic response of the electrode from DFT.
We found that the charge is a better measure of the maximum point compared to the potential because the charge directly determines the interfacial electric field seen by the water surface, while the potential depends more globally on the overall electrostatic potential.
Similarly, we expect the potential of maximum entropy (PME) relative to PZC to be inversely proportional to the peak capacitance for a family of ideal electrochemical interfaces with different capacitances, while the charge of maximum entropy (CME) will be  constant.
Consequently, we will focus on comparing the CME to the CMC predicted above.

\begin{figure}
\includegraphics[width=\columnwidth]{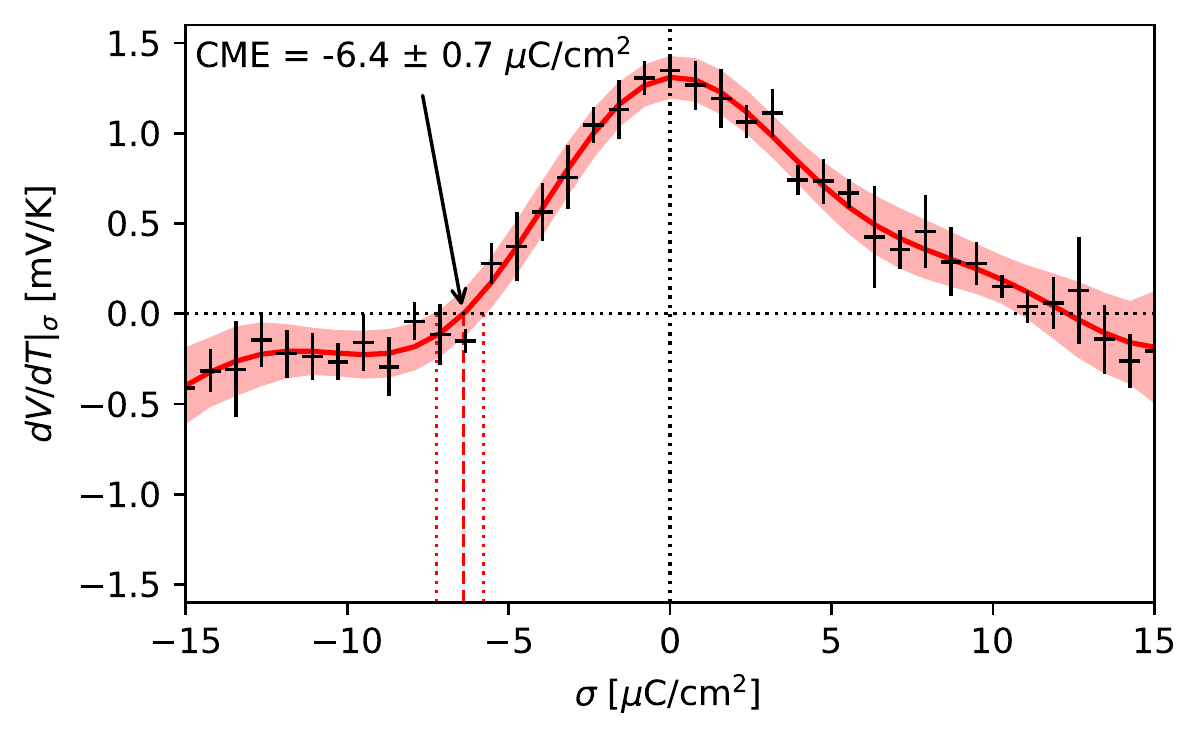}
\caption{Temperature derivative of interface potential, $dV/dT$, at several fixed electrode charges, $\sigma$.
The shaded region is a 95 \% confidence interval estimated using kernel ridge regression on several samplings of one of the five molecular dynamics runs at each charge density.
The zero crossing of $dV/dT|_\sigma = -dS/d\sigma|_T$ at $(-6.4 \pm 0.7)~\mu$C/cm$^2$ corresponds to the charge of maximum entropy (CME).}
\label{fig:dphi_dT}
\end{figure}

We predict CME by directly mimicking the experimental approach: heat the electrochemical interface and measure the change of electrode potential $\partial V/\partial T$ at fixed charge.
Specifically, we repeat the entire set of simulations used to generate the above results (which were for 298~K) at 318~K, and compute $\partial V/\partial T$ as a finite-difference derivative for each electrode charge density, $\sigma$ (FIG.~\ref{fig:dphi_dT}).
We find that $\partial V/\partial T$ crosses zero with a positive slope near $-6.4~\mu$C/cm$^2$, and since $\partial S/\partial \sigma|_T = -\partial V/\partial T|_\sigma$, this implies $\partial S/\partial \sigma$ will cross zero at this point with a negative slope.
Therefore, $\sigma \approx -6.4~\mu$C/cm$^2$ should be a local maximum of entropy as a function of electrode charge.

The error bars shown in FIG.~\ref{fig:dphi_dT} are calculated as the standard deviation of the $\partial V/\partial T$ calculated from each of the ten half-cells simulated for each charge.
Note the extremely small magnitude of the results: the potentials only change by approximately $10-20$~mV over our 20~K baseline; the CME prediction therefore requires large ensembles with sufficient statistics and a careful analysis.
To precisely pin down the zero-crossing of $\partial V/\partial T$, we fit a general Kernel ridge regression model so as to not bias the result by choosing a more restrictive function form such as a polynomial.
We repeat the fit for 1000 re-samplings of the data by selecting the result from different half-cells at each charge.
The band shown in FIG.~\ref{fig:dphi_dT} is the 90 \% confidence interval obtained from this ensemble of fits.
Finally, from the zero-crossing of this band, we can quantify the 90 \% confidence interval of the CME to be $(-6.4 \pm 0.7)~\mu$C/cm$^2$.

Our predicted CME for ideal metal-water interfaces is in excellent agreement with experimental measurements of $-5~\mu$C/cm$^2$ for gold\cite{laserPME} and ($-4$ to $-6$)$~\mu$C/cm$^2$ for mercury.\cite{HgPME}
Most interestingly, it differs from the capacitance peak location (CMC) of $-3.7~\mu$C/cm$^2$ to $-3.3~\mu$C/cm$^2$ from Section~\ref{sec:ResultsCapacitance}.
Also note that the 20~K difference used in the calculation of $\partial V/\partial T$ does not affect the interface properties appreciably and cannot be the reason for the difference between CMC and CME: the CMC shifts by at most $0.3~\mu$C/cm$^2$ between 298~K and 318~K (FIG.~\ref{fig:cap-318K} in Appendix~\ref{sec:AppendixResults}).
The CMC and CME indicate asymmetric charge and thermodynamic response in the same direction: higher for negatively-charged electrodes.
However, by carefully calculating both quantities from the same set of MD simulations, we can unambiguously conclude that the CMC and CME do not coincide even for ideal electrochemical interfaces.
This provides a renewed incentive to experimentally measure the CMC, which as discussed in the Introduction, is challenging because the low ionic concentrations typically used to avoid ion adsorption lead to a capacitance dip that obscures the precise location of the capacitance maximum.

\section*{Conclusions}
We have performed classical molecular dynamics simulations and evaluated the capacitance and potential of maximum entropy for aqueous, charged metallic interfaces.
We find distinct, non-coincident values for the CMC and CME.  For surfaces with large capacitance, the potentials of minimum entropy and maximum capacitance will be very similar.
Our findings of the asymmetric response of interfacial water open new questions about the electrochemical interface and the response properties themselves.

Future work is necessary to understand why the CME and CMC do not coincide even for ideal interfaces.
This could stem from the different spatial regions that contribute to each effect.
The entropy is sensitive to the entire polarized region of each half cell in the interface, while the the interfacial (series) capacitance is most sensitive to the lowest-capacitance region in space: closest to the metal.
These spatial dependencies could be explored further by varying the ionic concentration, and the generality of CME-CMC differences can be tested using MD simulations of other asymmetric solvents such as acetonitrile.\cite{CANDLE}

Going beyond ideal interfaces, extensions of this approach can  systematically quantify the impact of specific electrode and electrolyte properties on the charge response and thermodynamics of the double layer.
In particular, hydrophobicity of the electrode and ion sizes are known to impact the structure of interfacial water.\cite{NetzInterfacial, LimmerHydrophobic}
Lastly, the detailed capacitance and entropy predictions of charged interfaces presented here will facilitate future development of more accurate solvation models for electrochemistry.

\section*{Acknowledgements}
AS and RS acknowledge support by the U.S. Department of Energy, Office of Science, Basic Energy Sciences, under Award \#DE-SC0022247.
All calculations were carried out at the Center for Computational Innovations at Rensselaer Polytechnic Institute.
We acknowledge Dr. Thomas P. Moffat for his useful suggestions.

\appendix

\section{Force field and method details}\label{sec:AppendixMethods}

Section~\ref{sec:DetailsMD} describes the overall set of molecular dynamics (MD) simulations we carry out in the Large-scale Atomic/Molecular Massively Parallel Simulator, LAMMPS.\cite{LAMMPS}
Briefly, we carry out simulations of extended simple point charge model (SPC/E) water\cite{SPCE} with 1~mol/L NaF electrolyte\cite{Fyta2012} occupying a $45\times 45\times 43.6$~\Angstrom$^3$ volume between Ag(100) electrodes with various surface charge densities.
Here, we specify details of the force fields used in all the simulations discussed above.

\begin{table}[ht!]
\caption{Lennard-Jones coefficients for NaF electrolyte\cite{Fyta2012} in SPC/E water.\cite{SPCE}
The H atoms have zero LJ interactions in this model, and the atomic charges are $+0.4238$, $-0.8476$, $+1$ and $-1$ for H, O, Na$^+$ and F$^-$ respectively.
\label{tab:force-field}}
\begin{tabular}{|l l l|}
\hline
Atom/ion pair & $\epsilon$ (kcal/mol) & $\sigma$ (\Angstrom)\\
\hline
O-O & 0.1553 & 3.166\\
Na$^+$-O & 0.1247 & 2.876\\
Na$^+$-Na$^+$ & 0.1000 & 2.583\\
F$^-$-O & 0.02390 & 3.665\\
F$^-$-F$^-$ & 0.003585 & 4.161\\
Na$^+$-F$^-$ & 0.02840 & 3.372\\
\hline
\end{tabular}
\end{table}

The interactions within the electrolyte for SPC/E water and the ions are treated using short-ranged Lennard-Jones (LJ) and long-range Coulomb interactions, with the pair coefficients shown in Table~\ref{tab:force-field}.
Note that the LJ parameters follow Lorentz-Berthelot mixing of Lennard-Jones parameters (arithmetic for $\sigma$, geometric for $\epsilon$), except for between Na$^+$ and F$^-$, which is overridden for better description of ion interactions.\cite{Fyta2012}
The long-range Coulomb interaction is treated using the graphical processing unit (GPU) parallelized particle-particle particle-mesh (PPPM)\cite{pppm} method in LAMMPS, with relative accuracy of $10^{-5}$ in forces. 

\begin{table}
\caption{Morse potential parameters for electrode-electrolyte interactions. Note that Ag-Ne parameters are used for Ag-Na$^+$ and Ag-F$^-$.
\label{tab:morse}}
\begin{tabular}{|c|c|c|c|}
\hline
Atom Pair & $D_0$ (kcal/mol) & $\alpha$ (\Angstrom$^{-1}$) & $r_0$ (\Angstrom)\\
\hline
Ag-H  & 3.6425 & 0.8808 & 1.7453\\
Ag-O  & 1.2027 & 1.1462 & 2.9022\\
Ag-Ne & 2.1967 & 0.4426 & 5.2178\\
\hline
\end{tabular}
\end{table}

\begin{figure*}
\centering
\includegraphics[width=\textwidth]{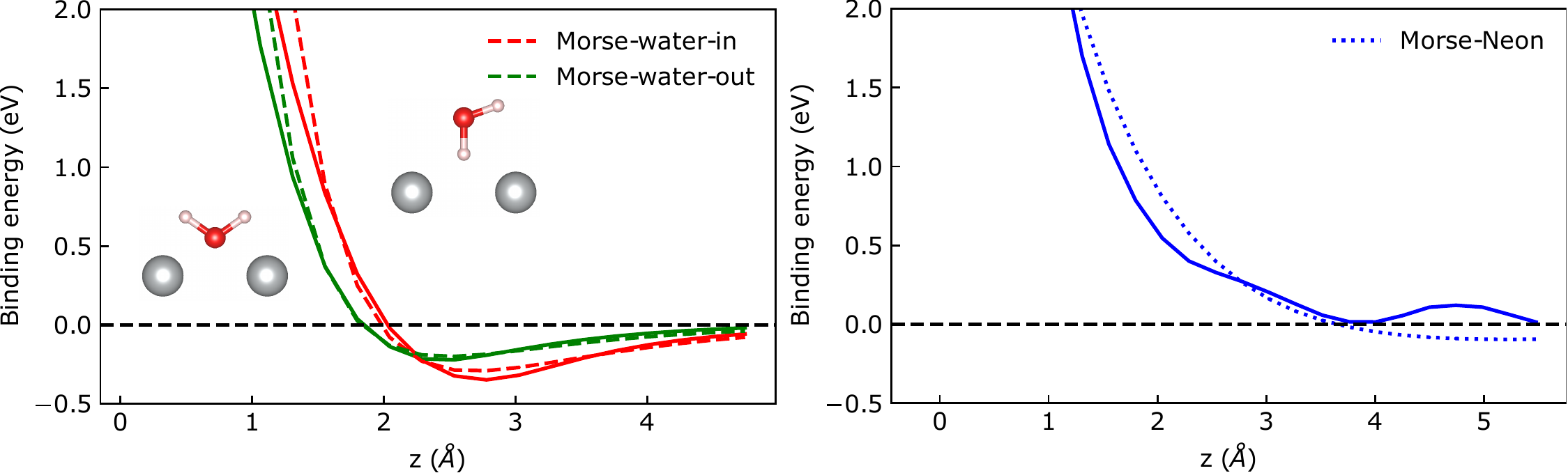}
\caption{Parameterization of the Morse potential between Ag(100) and the water (left) and ions (right).
The solid lines indicate the reference potential from DFT in each case, while the dashed / dotted lines are the best-fit Morse potentials.}
\label{fig:morse-fit}
\end{figure*}

\begin{figure}
\includegraphics[width=\columnwidth]{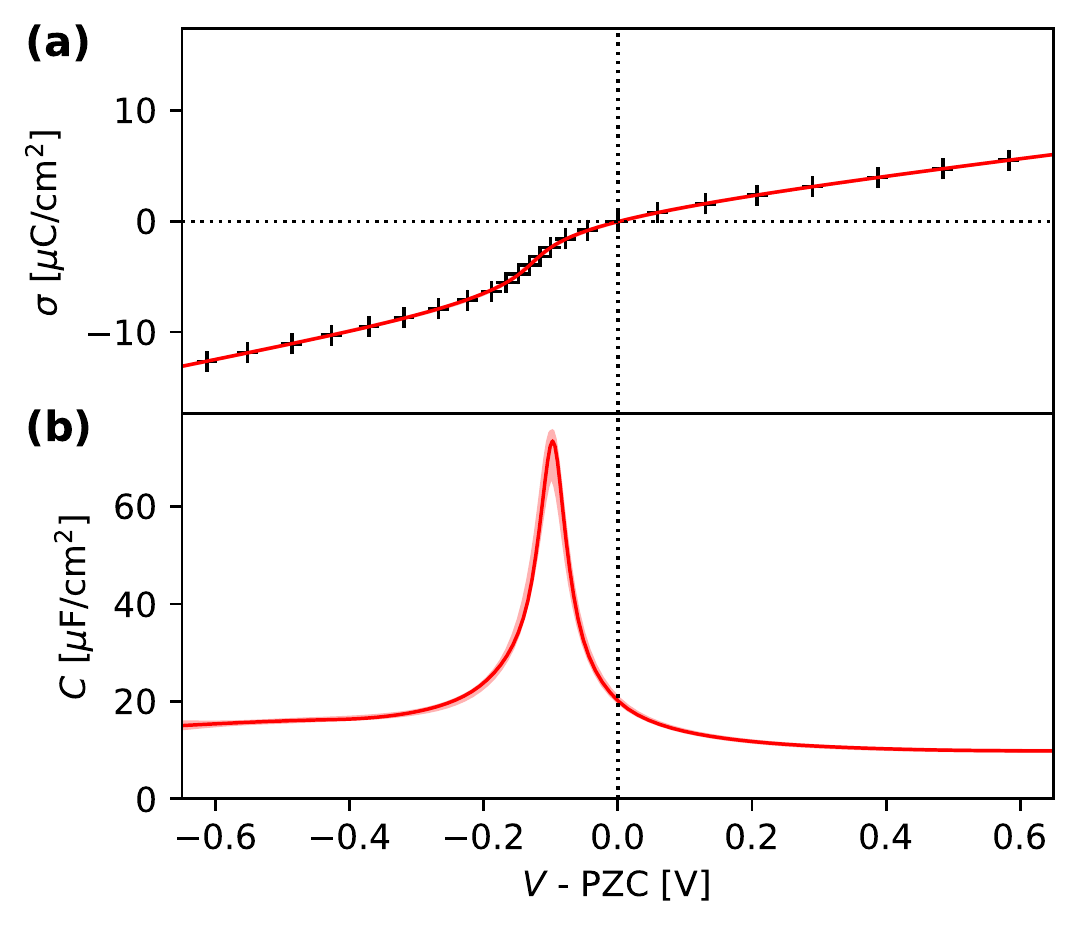}
\caption{Extraction of capacitance from MD potential ($V$) for each electrode charge density $\sigma$ (black +, shown here for the data corresponding to the $\Delta z = 0$ curve in FIG.~\ref{fig:capacitance}(a)).
(a) Charging curve and (b) corresponding capacitance from the best-fit piecewise cubic spline (red lines).
The pink region corresponds to a 95\% confidence interval estimated from an ensemble of splines fit to 200 random draws of 90\% of the data points.
Note that the confidence interval is invisible on the scale of the plot in (a), but visible in (b) due to amplification of numerical errors by the derivative computation for the capacitance.
From the ensemble, the maximum capacitance is at a charge, CMC = $(-3.7 \pm 0.2)~\mu$C/cm$^2$}
\label{fig:capExtraction}
\end{figure}

We use Morse potentials,
\begin{equation}
E(r) = D_0[e^{-2\alpha(r-r_0)}-2e^{-\alpha(r-r_0)}],
\end{equation}
to capture the interaction between the Ag atoms in the electrode and each atom type (H, O, Na$^+$ and F$^-$ ) in the electrolyte, with the parameters shown in Table~\ref{tab:morse}.
For the water, we calculate the density functional theory (DFT) binding energy of a water molecule as a function of distance away from a neutral Ag(100) surface, in two different configurations: both H pointing away and one H pointing towards the surface (FIG.~\ref{fig:morse-fit}).
We select the Morse potentials for Ag-O and Ag-H to best fit the interactions for both these orientations.
For the ions, we parameterize the interaction to a DFT binding energy curve of a neon atom to the silver surface since both Na$^+$ and F$^-$ are iso-electronic to neon and we primarily need to capture the repulsive part of the interaction to prevent ion escape from the electrolyte.
The parameters of the DFT calculations are as specified in Section~\ref{sec:DetailsDFT}, except that DFT-D2 corrections\cite{DFT-D2} are included to capture long-range dispersion interactions.

\begin{figure}
\includegraphics[width=\columnwidth]{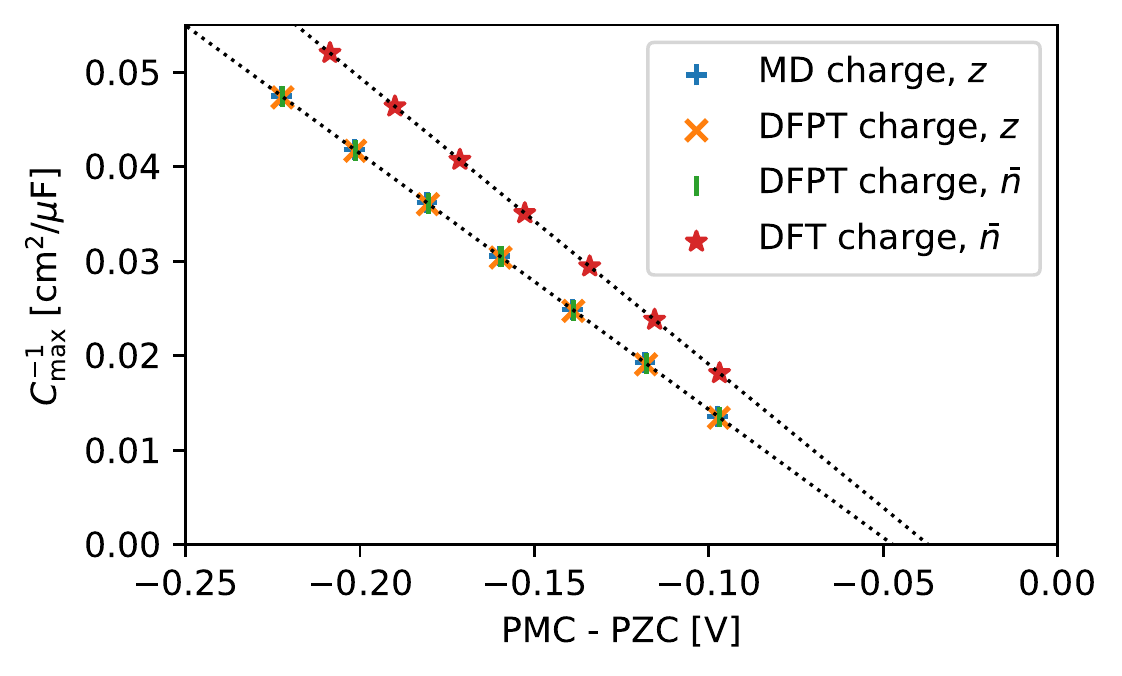}
\caption{Potential of maximum capacitance (PMC) relative to PZC is inversely proportional to the peak capacitance $C\sub{max}$ for the family of curves corresponding to each electrode charge model and placement scheme shown in FIG.~\ref{fig:capacitance}.
The dotted lines indicate ideal linear relation between $C\sub{max}^{-1}$ and PMC$-$PZC.}
\label{fig:capPeak}
\end{figure}

\begin{figure}
\centering
\includegraphics[width=\columnwidth]{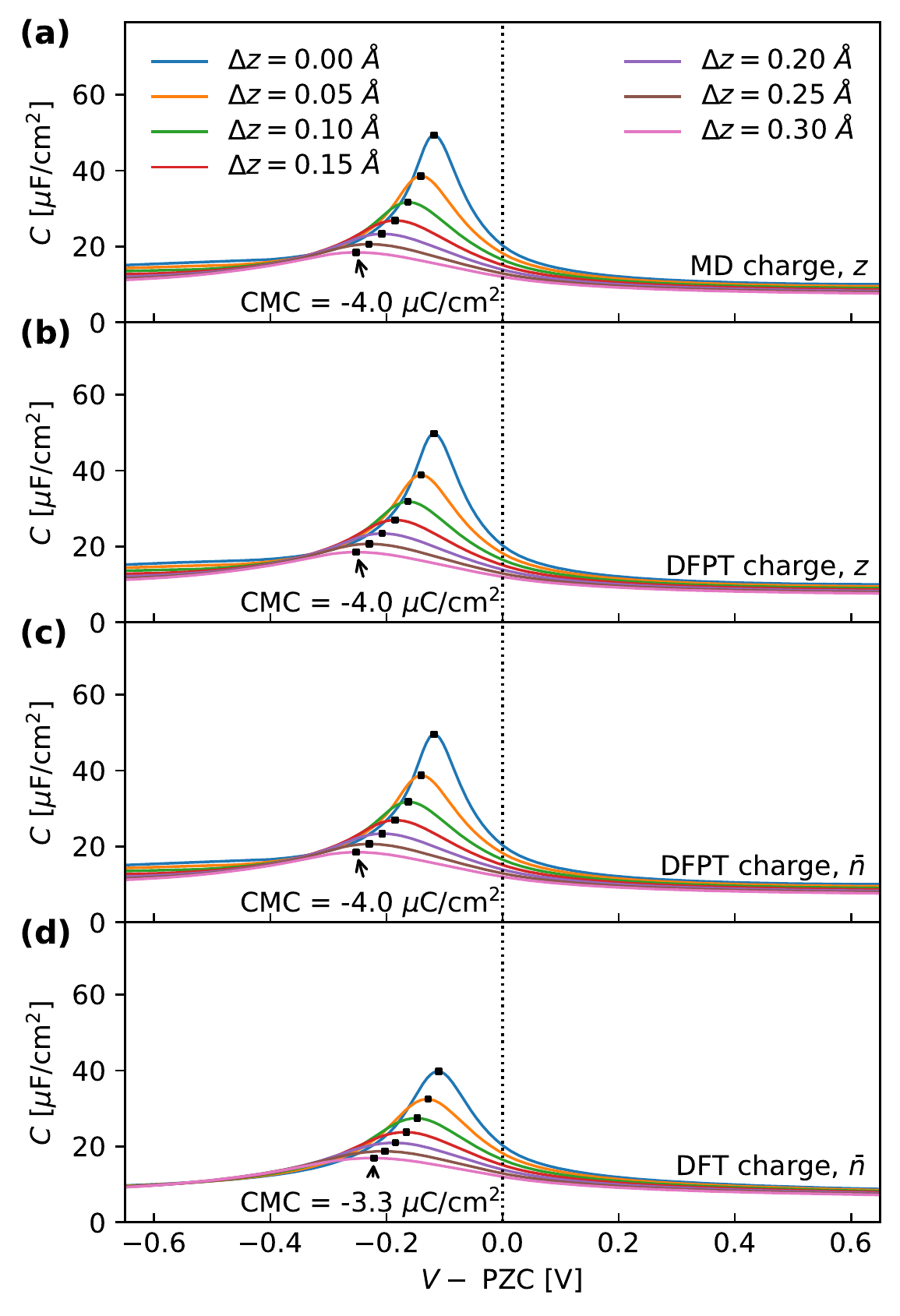}
\caption{Family of capacitance curves at 318~K, with different charge models and placement schemes analogous to the 298~K results shown in FIG.~\ref{fig:capacitance}.
The increase in temperature slightly changes the charge of maximum capacitance (CMC) from -3.7 to -4.0~$\mu$C/cm$^2$ in the first three panels, while the CMC remains unchanged in the fourth panel.}
\label{fig:cap-318K}
\end{figure}

FIG.~\ref{fig:capExtraction} shows the extraction of capacitance from the MD simulated potentials $V$ for each electrode charge $\sigma$.
We fit a piecewise cubic spline (using UnivariateSpline in SciPy \cite{SciPy}) to $V(\sigma)$ in order to compute the derivative $C_d^{-1} = \partial V/\partial\sigma$.
(We compute $\partial V/\partial\sigma$ because the points are evenly spaced in $\sigma$ but not in $V$.)
The spline has a single hyperparameter $s$ that sets the number of spline knots to the smallest value such that the sum of squared errors (SSE) is less than $s$.
We select this hyperparameter by maximizing the cross-validation score on a 5-fold split of the data, and find the optimum $s = 10^{-4}$~V$^2$.
To estimate error, we fit an ensemble of models to 200 distinct random draws of 90\% of the data and extract a 95\% confidence interval.
Numerical noise in the MD data is negligible, and the confidence interval is invisible for the charging curves (FIG.~\ref{fig:capExtraction}(a)).
However, the derivative of the spline is more noise sensitive and results in a visible confidence interval for the capacitance in FIG.~\ref{fig:capExtraction}(b).
Finally, we also locate the maximum of each capacitance curve in the ensemble to get statistics for the CMC, and find that the CMC is $(-3.7 \pm 0.2)~\mu$C/cm$^2$.

\section{Supplementary results}\label{sec:AppendixResults}

FIG.~\ref{fig:capPeak} shows the inverse proportionality between the potential of maximum capacitance (PMC) relative to PZC and the peak capacitance value for the families of capaictance curves generated with different offsets $\Delta z$ in the placement of the electrode and electrolyte charge densities.
This is expected because the charge of maximum capacitance is a constant across each family, as discussed in .
By definition, CMC = $\int\sub{PZC}\super{PMC} C_d dV = (\mathrm{PMC}-\mathrm{PZC}) \bar{C}$, where $\bar{C}$ is the average capacitance in the region between PZC and PMC.
Since the capacitance at the PZC and PMC are scaled similarly with changing $\Delta z$, we expect $\bar{C} \propto C_{max}$ and hence $C\sub{max}^{-1} \propto (\mathrm{PMC}-\mathrm{PZC})$.

FIG.~\ref{fig:cap-318K} shows the capacitance curves analogous to FIG.~\ref{fig:capacitance}, but computed from the MD simulations at 318~K instead of 298~K.
The peak in the capacitance curves broaden slightly with increased temperature, and the charge of maximum capacitance (CMC) changes by less than $0.3~\mu$C/cm$^2$.
This indicates that the 20~K difference in temperature used for the finite difference derivative in the estimation of the charge of maximum entropy (CME) does not substantially change the interfacial properties.
In particular, the $>2.7~\mu$C/cm$^2$ difference between CMC and CME is an order of magnitude greater than possible finite-difference errors from the 20~K difference.


\section*{References}
\begin{thebibliography}{59}\makeatletter
\providecommand \@ifxundefined [1]{\@ifx{#1\undefined}
}\providecommand \@ifnum [1]{\ifnum #1\expandafter \@firstoftwo
 \else \expandafter \@secondoftwo
 \fi
}\providecommand \@ifx [1]{\ifx #1\expandafter \@firstoftwo
 \else \expandafter \@secondoftwo
 \fi
}\providecommand \natexlab [1]{#1}\providecommand \enquote  [1]{``#1''}\providecommand \bibnamefont  [1]{#1}\providecommand \bibfnamefont [1]{#1}\providecommand \citenamefont [1]{#1}\providecommand \href@noop [0]{\@secondoftwo}\providecommand \href [0]{\begingroup \@sanitize@url \@href}\providecommand \@href[1]{\@@startlink{#1}\@@href}\providecommand \@@href[1]{\endgroup#1\@@endlink}\providecommand \@sanitize@url [0]{\catcode `\\12\catcode `\$12\catcode
  `\&12\catcode `\#12\catcode `\^12\catcode `\_12\catcode `\%12\relax}\providecommand \@@startlink[1]{}\providecommand \@@endlink[0]{}\providecommand \url  [0]{\begingroup\@sanitize@url \@url }\providecommand \@url [1]{\endgroup\@href {#1}{\urlprefix }}\providecommand \urlprefix  [0]{URL }\providecommand \Eprint [0]{\href }\providecommand \doibase [0]{https://doi.org/}\providecommand \selectlanguage [0]{\@gobble}\providecommand \bibinfo  [0]{\@secondoftwo}\providecommand \bibfield  [0]{\@secondoftwo}\providecommand \translation [1]{[#1]}\providecommand \BibitemOpen [0]{}\providecommand \bibitemStop [0]{}\providecommand \bibitemNoStop [0]{.\EOS\space}\providecommand \EOS [0]{\spacefactor3000\relax}\providecommand \BibitemShut  [1]{\csname bibitem#1\endcsname}\let\auto@bib@innerbib\@empty
\bibitem [{\citenamefont {Badwal}\ \emph {et~al.}(2014)\citenamefont {Badwal},
  \citenamefont {Giddey}, \citenamefont {Munnings}, \citenamefont {Bhatt},\
  and\ \citenamefont {Hollenkamp}}]{RevEnergyConversion}\BibitemOpen
  \bibfield  {author} {\bibinfo {author} {\bibfnamefont {S.~P.~S.}\
  \bibnamefont {Badwal}}, \bibinfo {author} {\bibfnamefont {S.~S.}\
  \bibnamefont {Giddey}}, \bibinfo {author} {\bibfnamefont {C.}~\bibnamefont
  {Munnings}}, \bibinfo {author} {\bibfnamefont {A.~I.}\ \bibnamefont
  {Bhatt}},\ and\ \bibinfo {author} {\bibfnamefont {A.~F.}\ \bibnamefont
  {Hollenkamp}},\ }\href {https://doi.org/10.3389/fchem.2014.00079} {\bibfield
  {journal} {\bibinfo  {journal} {Front. Chem.}\ }\textbf {\bibinfo {volume}
  {2}},\ \bibinfo {pages} {79} (\bibinfo {year} {2014})}\BibitemShut {NoStop}\bibitem [{\citenamefont {Gür}(2018)}]{RevEnergyStorage}\BibitemOpen
  \bibfield  {author} {\bibinfo {author} {\bibfnamefont {T.~M.}\ \bibnamefont
  {Gür}},\ }\href {https://doi.org/10.1039/C8EE01419A} {\bibfield  {journal}
  {\bibinfo  {journal} {Energy and Environmental Science}\ }\textbf {\bibinfo
  {volume} {11}},\ \bibinfo {pages} {2696} (\bibinfo {year}
  {2018})}\BibitemShut {NoStop}\bibitem [{\citenamefont {Goodenough}(2014)}]{Goodenough}\BibitemOpen
  \bibfield  {author} {\bibinfo {author} {\bibfnamefont {J.~B.}\ \bibnamefont
  {Goodenough}},\ }\href {https://doi.org/10.1039/C3EE42613K} {\bibfield
  {journal} {\bibinfo  {journal} {Energy Environ. Sci.}\ }\textbf {\bibinfo
  {volume} {7}},\ \bibinfo {pages} {14} (\bibinfo {year} {2014})}\BibitemShut
  {NoStop}\bibitem [{\citenamefont {Mçhle}\ \emph {et~al.}(2018)\citenamefont {Mçhle},
  \citenamefont {Zirbes}, \citenamefont {Rodrigo}, \citenamefont {Gieshoff},
  \citenamefont {Wiebe},\ and\ \citenamefont {Waldvogel}}]{RevChemSyn}\BibitemOpen
  \bibfield  {author} {\bibinfo {author} {\bibfnamefont {S.}~\bibnamefont
  {Mçhle}}, \bibinfo {author} {\bibfnamefont {M.}~\bibnamefont {Zirbes}},
  \bibinfo {author} {\bibfnamefont {E.}~\bibnamefont {Rodrigo}}, \bibinfo
  {author} {\bibfnamefont {T.}~\bibnamefont {Gieshoff}}, \bibinfo {author}
  {\bibfnamefont {A.}~\bibnamefont {Wiebe}},\ and\ \bibinfo {author}
  {\bibfnamefont {S.~R.}\ \bibnamefont {Waldvogel}},\ }\href@noop {} {\bibfield
   {journal} {\bibinfo  {journal} {Angew. Chem. Int. Ed.}\ }\textbf {\bibinfo
  {volume} {57}},\ \bibinfo {pages} {6018} (\bibinfo {year}
  {2018})}\BibitemShut {NoStop}\bibitem [{\citenamefont {Climent}\ \emph
  {et~al.}(2002{\natexlab{a}})\citenamefont {Climent}, \citenamefont {Coles},\
  and\ \citenamefont {Compton}}]{Pt111PME2}\BibitemOpen
  \bibfield  {author} {\bibinfo {author} {\bibfnamefont {V.}~\bibnamefont
  {Climent}}, \bibinfo {author} {\bibfnamefont {B.~A.}\ \bibnamefont {Coles}},\
  and\ \bibinfo {author} {\bibfnamefont {R.~G.}\ \bibnamefont {Compton}},\
  }\href {https://doi.org/https://doi.org/10.1021/jp020785q} {\bibfield
  {journal} {\bibinfo  {journal} {J. Phys. Chem. B}\ }\textbf {\bibinfo
  {volume} {106}},\ \bibinfo {pages} {5988–5996} (\bibinfo {year}
  {2002}{\natexlab{a}})}\BibitemShut {NoStop}\bibitem [{\citenamefont {Harrison}\ \emph {et~al.}(1973)\citenamefont
  {Harrison}, \citenamefont {Randles},\ and\ \citenamefont
  {Schiffrin}}]{HgPME}\BibitemOpen
  \bibfield  {author} {\bibinfo {author} {\bibfnamefont {J.~A.}\ \bibnamefont
  {Harrison}}, \bibinfo {author} {\bibfnamefont {J.~E.~B.}\ \bibnamefont
  {Randles}},\ and\ \bibinfo {author} {\bibfnamefont {D.~J.}\ \bibnamefont
  {Schiffrin}},\ }\href@noop {} {\bibfield  {journal} {\bibinfo  {journal}
  {Electroanalytical Chemistry and Interfacial Electrochemistry}\ }\textbf
  {\bibinfo {volume} {4}},\ \bibinfo {pages} {359} (\bibinfo {year}
  {1973})}\BibitemShut {NoStop}\bibitem [{\citenamefont {Climent}\ \emph
  {et~al.}(2002{\natexlab{b}})\citenamefont {Climent}, \citenamefont {Coles},\
  and\ \citenamefont {Compton}}]{laserPME}\BibitemOpen
  \bibfield  {author} {\bibinfo {author} {\bibfnamefont {V.}~\bibnamefont
  {Climent}}, \bibinfo {author} {\bibfnamefont {B.~A.}\ \bibnamefont {Coles}},\
  and\ \bibinfo {author} {\bibfnamefont {R.~G.}\ \bibnamefont {Compton}},\
  }\href {https://doi.org/https://doi.org/10.1021/jp020054q} {\bibfield
  {journal} {\bibinfo  {journal} {Phys. Chem. B}\ }\textbf {\bibinfo {volume}
  {106}},\ \bibinfo {pages} {5258} (\bibinfo {year}
  {2002}{\natexlab{b}})}\BibitemShut {NoStop}\bibitem [{\citenamefont {Ding}\ \emph {et~al.}(2021)\citenamefont {Ding},
  \citenamefont {Garlyyev}, \citenamefont {Watzele}, \citenamefont {Sarpey},\
  and\ \citenamefont {Bandarenka}}]{PMEElectrolytes}\BibitemOpen
  \bibfield  {author} {\bibinfo {author} {\bibfnamefont {X.}~\bibnamefont
  {Ding}}, \bibinfo {author} {\bibfnamefont {B.}~\bibnamefont {Garlyyev}},
  \bibinfo {author} {\bibfnamefont {S.~A.}\ \bibnamefont {Watzele}}, \bibinfo
  {author} {\bibfnamefont {T.~K.}\ \bibnamefont {Sarpey}},\ and\ \bibinfo
  {author} {\bibfnamefont {A.~S.}\ \bibnamefont {Bandarenka}},\ }\href
  {https://doi.org/https://doi.org/10.1002/chem.202101537} {\bibfield
  {journal} {\bibinfo  {journal} {Chem. Eur. J.}\ }\textbf {\bibinfo {volume}
  {27}},\ \bibinfo {pages} {10016} (\bibinfo {year} {2021})}\BibitemShut
  {NoStop}\bibitem [{\citenamefont {Sarabia}\ \emph {et~al.}(2019)\citenamefont
  {Sarabia}, \citenamefont {Sebastián-Pascual}, \citenamefont {Koper},
  \citenamefont {Climent},\ and\ \citenamefont {Feliu}}]{KoperNiOH}\BibitemOpen
  \bibfield  {author} {\bibinfo {author} {\bibfnamefont {F.~J.}\ \bibnamefont
  {Sarabia}}, \bibinfo {author} {\bibfnamefont {P.}~\bibnamefont
  {Sebastián-Pascual}}, \bibinfo {author} {\bibfnamefont {M.~T.}\ \bibnamefont
  {Koper}}, \bibinfo {author} {\bibfnamefont {V.}~\bibnamefont {Climent}},\
  and\ \bibinfo {author} {\bibfnamefont {J.~M.}\ \bibnamefont {Feliu}},\ }\href
  {https://doi.org/https://doi.org/10.1021/acsami.8b15003} {\bibfield
  {journal} {\bibinfo  {journal} {ACS Appl. Mater. Interfaces}\ }\textbf
  {\bibinfo {volume} {11}},\ \bibinfo {pages} {613–623} (\bibinfo {year}
  {2019})}\BibitemShut {NoStop}\bibitem [{\citenamefont {Climent}\ \emph {et~al.}(2006)\citenamefont
  {Climent}, \citenamefont {Garcia-Araez}, \citenamefont {Compton},\ and\
  \citenamefont {Feliu}}]{PtBiPME}\BibitemOpen
  \bibfield  {author} {\bibinfo {author} {\bibfnamefont {V.}~\bibnamefont
  {Climent}}, \bibinfo {author} {\bibfnamefont {N.}~\bibnamefont
  {Garcia-Araez}}, \bibinfo {author} {\bibfnamefont {R.~G.}\ \bibnamefont
  {Compton}},\ and\ \bibinfo {author} {\bibfnamefont {J.~M.}\ \bibnamefont
  {Feliu}},\ }\href {https://doi.org/https://doi.org/10.1021/jp061982i}
  {\bibfield  {journal} {\bibinfo  {journal} {J. Phys. Chem. B}\ }\textbf
  {\bibinfo {volume} {110}},\ \bibinfo {pages} {21092–21100} (\bibinfo {year}
  {2006})}\BibitemShut {NoStop}\bibitem [{\citenamefont {Martínez-Hincapié}\ \emph
  {et~al.}(2017)\citenamefont {Martínez-Hincapié}, \citenamefont
  {Sebastián-Pascual}, \citenamefont {Climent},\ and\ \citenamefont
  {Feliu}}]{FeliuPME}\BibitemOpen
  \bibfield  {author} {\bibinfo {author} {\bibfnamefont {R.}~\bibnamefont
  {Martínez-Hincapié}}, \bibinfo {author} {\bibfnamefont {P.}~\bibnamefont
  {Sebastián-Pascual}}, \bibinfo {author} {\bibfnamefont {V.}~\bibnamefont
  {Climent}},\ and\ \bibinfo {author} {\bibfnamefont {J.~M.}\ \bibnamefont
  {Feliu}},\ }\href@noop {} {\bibfield  {journal} {\bibinfo  {journal} {Russian
  Journal of Electrochemistry}\ ,\ \bibinfo {pages} {227}} (\bibinfo {year}
  {2017})}\BibitemShut {NoStop}\bibitem [{\citenamefont {Sebastián}\ \emph {et~al.}(2017)\citenamefont
  {Sebastián}, \citenamefont {Martínez-Hincapié}, \citenamefont {Climent},\
  and\ \citenamefont {Feliu}}]{FeliuPMEPt}\BibitemOpen
  \bibfield  {author} {\bibinfo {author} {\bibfnamefont {P.}~\bibnamefont
  {Sebastián}}, \bibinfo {author} {\bibfnamefont {R.}~\bibnamefont
  {Martínez-Hincapié}}, \bibinfo {author} {\bibfnamefont {V.}~\bibnamefont
  {Climent}},\ and\ \bibinfo {author} {\bibfnamefont {J.}~\bibnamefont
  {Feliu}},\ }\href
  {https://doi.org/https://doi.org/10.1016/j.electacta.2017.01.089} {\bibfield
  {journal} {\bibinfo  {journal} {Electrochimica Acta}\ }\textbf {\bibinfo
  {volume} {228}},\ \bibinfo {pages} {667} (\bibinfo {year}
  {2017})}\BibitemShut {NoStop}\bibitem [{\citenamefont {Ganassin}\ \emph {et~al.}(2017)\citenamefont
  {Ganassin}, \citenamefont {Sebastián}, \citenamefont {Climent},
  \citenamefont {Schuhmann}, \citenamefont {Bandarenka},\ and\ \citenamefont
  {Feliu}}]{IrPME}\BibitemOpen
  \bibfield  {author} {\bibinfo {author} {\bibfnamefont {A.}~\bibnamefont
  {Ganassin}}, \bibinfo {author} {\bibfnamefont {P.}~\bibnamefont
  {Sebastián}}, \bibinfo {author} {\bibfnamefont {V.}~\bibnamefont {Climent}},
  \bibinfo {author} {\bibfnamefont {W.}~\bibnamefont {Schuhmann}}, \bibinfo
  {author} {\bibfnamefont {A.~S.}\ \bibnamefont {Bandarenka}},\ and\ \bibinfo
  {author} {\bibfnamefont {J.}~\bibnamefont {Feliu}},\ }\href
  {https://doi.org/https://doi.org/10.1038/s41598-017-01295-1} {\bibfield
  {journal} {\bibinfo  {journal} {Sci Rep}\ }\textbf {\bibinfo {volume} {7}},\
  \bibinfo {pages} {1246} (\bibinfo {year} {2017})}\BibitemShut {NoStop}\bibitem [{\citenamefont {Sebastiań-Pascual}\ \emph
  {et~al.}(2020)\citenamefont {Sebastiań-Pascual}, \citenamefont {Sarabia},
  \citenamefont {Climent}, \citenamefont {Feliu},\ and\ \citenamefont
  {Escudero-Escribano}}]{CuPME}\BibitemOpen
  \bibfield  {author} {\bibinfo {author} {\bibfnamefont {P.}~\bibnamefont
  {Sebastiań-Pascual}}, \bibinfo {author} {\bibfnamefont {F.~J.}\ \bibnamefont
  {Sarabia}}, \bibinfo {author} {\bibfnamefont {V.}~\bibnamefont {Climent}},
  \bibinfo {author} {\bibfnamefont {J.~M.}\ \bibnamefont {Feliu}},\ and\
  \bibinfo {author} {\bibfnamefont {M.}~\bibnamefont {Escudero-Escribano}},\
  }\href {https://doi.org/https://dx.doi.org/10.1021/acs.jpcc.0c07821}
  {\bibfield  {journal} {\bibinfo  {journal} {J. Phys. Chem. C}\ }\textbf
  {\bibinfo {volume} {124}},\ \bibinfo {pages} {23253} (\bibinfo {year}
  {2020})}\BibitemShut {NoStop}\bibitem [{\citenamefont {Montenegro}\ \emph {et~al.}(2021)\citenamefont
  {Montenegro}, \citenamefont {Dutta}, \citenamefont {Mammetkuliev},
  \citenamefont {Shi}, \citenamefont {Hou}, \citenamefont {Bhattacharyya},
  \citenamefont {Zhao}, \citenamefont {Cronin},\ and\ \citenamefont
  {Benderskii}}]{AsymmWater}\BibitemOpen
  \bibfield  {author} {\bibinfo {author} {\bibfnamefont {A.}~\bibnamefont
  {Montenegro}}, \bibinfo {author} {\bibfnamefont {C.}~\bibnamefont {Dutta}},
  \bibinfo {author} {\bibfnamefont {M.}~\bibnamefont {Mammetkuliev}}, \bibinfo
  {author} {\bibfnamefont {H.}~\bibnamefont {Shi}}, \bibinfo {author}
  {\bibfnamefont {B.}~\bibnamefont {Hou}}, \bibinfo {author} {\bibfnamefont
  {D.}~\bibnamefont {Bhattacharyya}}, \bibinfo {author} {\bibfnamefont
  {B.}~\bibnamefont {Zhao}}, \bibinfo {author} {\bibfnamefont {S.~B.}\
  \bibnamefont {Cronin}},\ and\ \bibinfo {author} {\bibfnamefont {A.~V.}\
  \bibnamefont {Benderskii}},\ }\href
  {https://doi.org/https://doi.org/10.1038/s41586-021-03504-4} {\bibfield
  {journal} {\bibinfo  {journal} {Nature}\ }\textbf {\bibinfo {volume} {594}},\
  \bibinfo {pages} {62–65} (\bibinfo {year} {2021})}\BibitemShut {NoStop}\bibitem [{\citenamefont {Zhang}\ \emph {et~al.}(2020)\citenamefont {Zhang},
  \citenamefont {Stirnemann}, \citenamefont {Hynes},\ and\ \citenamefont
  {Laage}}]{Dynamics}\BibitemOpen
  \bibfield  {author} {\bibinfo {author} {\bibfnamefont {Y.}~\bibnamefont
  {Zhang}}, \bibinfo {author} {\bibfnamefont {G.}~\bibnamefont {Stirnemann}},
  \bibinfo {author} {\bibfnamefont {J.~T.}\ \bibnamefont {Hynes}},\ and\
  \bibinfo {author} {\bibfnamefont {D.}~\bibnamefont {Laage}},\ }\href
  {https://pubs.rsc.org/en/content/articlelanding/2020/CP/D0CP00359J#fn1}
  {\bibfield  {journal} {\bibinfo  {journal} {Phys. Chem. Chem. Phys.}\
  }\textbf {\bibinfo {volume} {22}},\ \bibinfo {pages} {10581} (\bibinfo {year}
  {2020})}\BibitemShut {NoStop}\bibitem [{\citenamefont {Sundararaman}\ \emph {et~al.}(2014)\citenamefont
  {Sundararaman}, \citenamefont {Letchworth-Weaver},\ and\ \citenamefont
  {Arias}}]{PolarizableCDFT}\BibitemOpen
  \bibfield  {author} {\bibinfo {author} {\bibfnamefont {R.}~\bibnamefont
  {Sundararaman}}, \bibinfo {author} {\bibfnamefont {K.}~\bibnamefont
  {Letchworth-Weaver}},\ and\ \bibinfo {author} {\bibfnamefont {T.~A.}\
  \bibnamefont {Arias}},\ }\href@noop {} {\bibfield  {journal} {\bibinfo
  {journal} {J . Chem. Phys.}\ }\textbf {\bibinfo {volume} {140}},\ \bibinfo
  {pages} {144504} (\bibinfo {year} {2014})}\BibitemShut {NoStop}\bibitem [{\citenamefont {Sundararaman}\ and\ \citenamefont
  {Goddard}(2015)}]{CANDLE}\BibitemOpen
  \bibfield  {author} {\bibinfo {author} {\bibfnamefont {R.}~\bibnamefont
  {Sundararaman}}\ and\ \bibinfo {author} {\bibfnamefont {W.~A.}\ \bibnamefont
  {Goddard}},\ }\bibfield  {journal} {\bibinfo  {journal} {J. Chem. Phys.}\
  }\textbf {\bibinfo {volume} {142}},\ \href
  {https://doi.org/10.1063/1.4907731} {10.1063/1.4907731} (\bibinfo {year}
  {2015})\BibitemShut {NoStop}\bibitem [{\citenamefont {Hu}\ \emph {et~al.}(2013)\citenamefont {Hu},
  \citenamefont {Vatamanu}, \citenamefont {Borodin},\ and\ \citenamefont
  {Bedrov}}]{NPTEchem}\BibitemOpen
  \bibfield  {author} {\bibinfo {author} {\bibfnamefont {Z.}~\bibnamefont
  {Hu}}, \bibinfo {author} {\bibfnamefont {J.}~\bibnamefont {Vatamanu}},
  \bibinfo {author} {\bibfnamefont {O.}~\bibnamefont {Borodin}},\ and\ \bibinfo
  {author} {\bibfnamefont {D.}~\bibnamefont {Bedrov}},\ }\href@noop {}
  {\bibfield  {journal} {\bibinfo  {journal} {Phys. Chem. Chem. Phys.}\
  }\textbf {\bibinfo {volume} {15}},\ \bibinfo {pages} {14234} (\bibinfo {year}
  {2013})}\BibitemShut {NoStop}\bibitem [{\citenamefont {Jiang}\ \emph {et~al.}(2016)\citenamefont {Jiang},
  \citenamefont {Cheng}, \citenamefont {Li},\ and\ \citenamefont
  {Liu}}]{GrapheneAqueousElectrolyte}\BibitemOpen
  \bibfield  {author} {\bibinfo {author} {\bibfnamefont {G.}~\bibnamefont
  {Jiang}}, \bibinfo {author} {\bibfnamefont {C.}~\bibnamefont {Cheng}},
  \bibinfo {author} {\bibfnamefont {D.}~\bibnamefont {Li}},\ and\ \bibinfo
  {author} {\bibfnamefont {J.~Z.}\ \bibnamefont {Liu}},\ }\href
  {https://doi.org/10.1007/s12274-015-0978-5} {\bibfield  {journal} {\bibinfo
  {journal} {Nano Research}\ }\textbf {\bibinfo {volume} {9}},\ \bibinfo
  {pages} {174–186} (\bibinfo {year} {2016})}\BibitemShut {NoStop}\bibitem [{\citenamefont {Demir}\ and\ \citenamefont
  {Searles}(2020)}]{IonicLiquidDLformation}\BibitemOpen
  \bibfield  {author} {\bibinfo {author} {\bibfnamefont {B.}~\bibnamefont
  {Demir}}\ and\ \bibinfo {author} {\bibfnamefont {D.~J.}\ \bibnamefont
  {Searles}},\ }\href {https://doi.org/10.3390/nano10112181} {\bibfield
  {journal} {\bibinfo  {journal} {Nanomaterials}\ }\textbf {\bibinfo {volume}
  {10}},\ \bibinfo {pages} {2181} (\bibinfo {year} {2020})}\BibitemShut
  {NoStop}\bibitem [{\citenamefont {Uralcan}\ \emph {et~al.}(2016)\citenamefont
  {Uralcan}, \citenamefont {Aksay}, \citenamefont {Debenedetti},\ and\
  \citenamefont {Limmer}}]{ChargeFluctIons}\BibitemOpen
  \bibfield  {author} {\bibinfo {author} {\bibfnamefont {B.}~\bibnamefont
  {Uralcan}}, \bibinfo {author} {\bibfnamefont {I.~A.}\ \bibnamefont {Aksay}},
  \bibinfo {author} {\bibfnamefont {P.~G.}\ \bibnamefont {Debenedetti}},\ and\
  \bibinfo {author} {\bibfnamefont {D.~T.}\ \bibnamefont {Limmer}},\
  }\href@noop {} {\bibfield  {journal} {\bibinfo  {journal} {J. Phys. Chem.
  Lett.}\ }\textbf {\bibinfo {volume} {7}},\ \bibinfo {pages} {2333–2338}
  (\bibinfo {year} {2016})}\BibitemShut {NoStop}\bibitem [{\citenamefont {Scalfi}\ \emph
  {et~al.}(2020{\natexlab{a}})\citenamefont {Scalfi}, \citenamefont {Limmer},
  \citenamefont {Coretti}, \citenamefont {Bonella}, \citenamefont {Madden},
  \citenamefont {Salanne},\ and\ \citenamefont
  {Rotenberg}}]{ChargeFluctElectrons}\BibitemOpen
  \bibfield  {author} {\bibinfo {author} {\bibfnamefont {L.}~\bibnamefont
  {Scalfi}}, \bibinfo {author} {\bibfnamefont {D.~T.}\ \bibnamefont {Limmer}},
  \bibinfo {author} {\bibfnamefont {A.}~\bibnamefont {Coretti}}, \bibinfo
  {author} {\bibfnamefont {S.}~\bibnamefont {Bonella}}, \bibinfo {author}
  {\bibfnamefont {P.~A.}\ \bibnamefont {Madden}}, \bibinfo {author}
  {\bibfnamefont {M.}~\bibnamefont {Salanne}},\ and\ \bibinfo {author}
  {\bibfnamefont {B.}~\bibnamefont {Rotenberg}},\ }\href@noop {} {\bibfield
  {journal} {\bibinfo  {journal} {Phys. Chem. Chem. Phys.}\ }\textbf {\bibinfo
  {volume} {22}},\ \bibinfo {pages} {10480} (\bibinfo {year}
  {2020}{\natexlab{a}})}\BibitemShut {NoStop}\bibitem [{\citenamefont {Limmer}\ \emph
  {et~al.}(2013{\natexlab{a}})\citenamefont {Limmer}, \citenamefont {Merle},
  \citenamefont {Salanne}, \citenamefont {Chandler}, \citenamefont {Madden},
  \citenamefont {van Roij},\ and\ \citenamefont {Rotenberg}}]{Chandler}\BibitemOpen
  \bibfield  {author} {\bibinfo {author} {\bibfnamefont {D.~T.}\ \bibnamefont
  {Limmer}}, \bibinfo {author} {\bibfnamefont {C.}~\bibnamefont {Merle}},
  \bibinfo {author} {\bibfnamefont {M.}~\bibnamefont {Salanne}}, \bibinfo
  {author} {\bibfnamefont {D.}~\bibnamefont {Chandler}}, \bibinfo {author}
  {\bibfnamefont {P.~A.}\ \bibnamefont {Madden}}, \bibinfo {author}
  {\bibfnamefont {R.}~\bibnamefont {van Roij}},\ and\ \bibinfo {author}
  {\bibfnamefont {B.}~\bibnamefont {Rotenberg}},\ }\href
  {https://doi.org/10.1103/PhysRevLett.111.106102} {\bibfield  {journal}
  {\bibinfo  {journal} {Phys. Rev. Lett.}\ }\textbf {\bibinfo {volume} {111}},\
  \bibinfo {pages} {106102} (\bibinfo {year} {2013}{\natexlab{a}})}\BibitemShut
  {NoStop}\bibitem [{\citenamefont {Voroshylova}\ \emph {et~al.}(2021)\citenamefont
  {Voroshylova}, \citenamefont {Ers}, \citenamefont {Koverga}, \citenamefont
  {Docampo-Álvarez}, \citenamefont {Pikma}, \citenamefont {Ivaništšev},\
  and\ \citenamefont {Cordeiro}}]{CapacitanceIssues1}\BibitemOpen
  \bibfield  {author} {\bibinfo {author} {\bibfnamefont {I.~V.}\ \bibnamefont
  {Voroshylova}}, \bibinfo {author} {\bibfnamefont {H.}~\bibnamefont {Ers}},
  \bibinfo {author} {\bibfnamefont {V.}~\bibnamefont {Koverga}}, \bibinfo
  {author} {\bibfnamefont {B.}~\bibnamefont {Docampo-Álvarez}}, \bibinfo
  {author} {\bibfnamefont {P.}~\bibnamefont {Pikma}}, \bibinfo {author}
  {\bibfnamefont {V.~B.}\ \bibnamefont {Ivaništšev}},\ and\ \bibinfo {author}
  {\bibfnamefont {M.~N.~D.}\ \bibnamefont {Cordeiro}},\ }\href
  {https://doi.org/https://doi.org/10.1016/j.electacta.2021.138148} {\bibfield
  {journal} {\bibinfo  {journal} {Electrochimica Acta}\ }\textbf {\bibinfo
  {volume} {379}},\ \bibinfo {pages} {138148} (\bibinfo {year}
  {2021})}\BibitemShut {NoStop}\bibitem [{\citenamefont {Haskins}\ and\ \citenamefont
  {Lawson}(2016)}]{CapacitanceIssues2}\BibitemOpen
  \bibfield  {author} {\bibinfo {author} {\bibfnamefont {J.~B.}\ \bibnamefont
  {Haskins}}\ and\ \bibinfo {author} {\bibfnamefont {J.~W.}\ \bibnamefont
  {Lawson}},\ }\href {https://doi.org/10.1063/1.4948938} {\bibfield  {journal}
  {\bibinfo  {journal} {The Journal of Chemical Physics}\ }\textbf {\bibinfo
  {volume} {144}},\ \bibinfo {pages} {184707} (\bibinfo {year}
  {2016})}\BibitemShut {NoStop}\bibitem [{\citenamefont {Bockris}\ \emph {et~al.}(2001)\citenamefont
  {Bockris}, \citenamefont {Reddy},\ and\ \citenamefont
  {Gamboa-Aldeco}}]{Bokris}\BibitemOpen
  \bibfield  {author} {\bibinfo {author} {\bibfnamefont {J.~O.}\ \bibnamefont
  {Bockris}}, \bibinfo {author} {\bibfnamefont {A.~K.}\ \bibnamefont {Reddy}},\
  and\ \bibinfo {author} {\bibfnamefont {M.~E.}\ \bibnamefont
  {Gamboa-Aldeco}},\ }\href@noop {} {\emph {\bibinfo {title} {Modern
  Electrochemistry 2A: Fundamentals of Electrodics}}}\ (\bibinfo  {publisher}
  {Springer Science \& Business Media},\ \bibinfo {year} {2001})\BibitemShut
  {NoStop}\bibitem [{\citenamefont {Ers}\ \emph {et~al.}(2020)\citenamefont {Ers},
  \citenamefont {Lembinen}, \citenamefont {Mišin}, \citenamefont {Seitsonen},
  \citenamefont {Fedorov},\ and\ \citenamefont {Ivaništšev}}]{PotDrop}\BibitemOpen
  \bibfield  {author} {\bibinfo {author} {\bibfnamefont {H.}~\bibnamefont
  {Ers}}, \bibinfo {author} {\bibfnamefont {M.}~\bibnamefont {Lembinen}},
  \bibinfo {author} {\bibfnamefont {M.}~\bibnamefont {Mišin}}, \bibinfo
  {author} {\bibfnamefont {A.~P.}\ \bibnamefont {Seitsonen}}, \bibinfo {author}
  {\bibfnamefont {M.~V.}\ \bibnamefont {Fedorov}},\ and\ \bibinfo {author}
  {\bibfnamefont {V.~B.}\ \bibnamefont {Ivaništšev}},\ }\href
  {https://doi.org/https://doi.org/10.1021/acs.jpcc.0c02964} {\bibfield
  {journal} {\bibinfo  {journal} {J. Phys. Chem. C}\ }\textbf {\bibinfo
  {volume} {124}},\ \bibinfo {pages} {19548–19555} (\bibinfo {year}
  {2020})}\BibitemShut {NoStop}\bibitem [{\citenamefont {Scalfi}\ \emph
  {et~al.}(2020{\natexlab{b}})\citenamefont {Scalfi}, \citenamefont {Dufils},
  \citenamefont {Reeves}, \citenamefont {Rotenberg},\ and\ \citenamefont
  {Salanne}}]{ThomasFermiMetallic}\BibitemOpen
  \bibfield  {author} {\bibinfo {author} {\bibfnamefont {L.}~\bibnamefont
  {Scalfi}}, \bibinfo {author} {\bibfnamefont {T.}~\bibnamefont {Dufils}},
  \bibinfo {author} {\bibfnamefont {K.~G.}\ \bibnamefont {Reeves}}, \bibinfo
  {author} {\bibfnamefont {B.}~\bibnamefont {Rotenberg}},\ and\ \bibinfo
  {author} {\bibfnamefont {M.}~\bibnamefont {Salanne}},\ }\href
  {https://doi.org/https://doi.org/10.1063/5.0028232} {\bibfield  {journal}
  {\bibinfo  {journal} {Journal of Chemical Physics}\ }\textbf {\bibinfo
  {volume} {153}},\ \bibinfo {pages} {174704} (\bibinfo {year}
  {2020}{\natexlab{b}})}\BibitemShut {NoStop}\bibitem [{\citenamefont {Kornyshev}\ \emph {et~al.}(2014)\citenamefont
  {Kornyshev}, \citenamefont {Luque},\ and\ \citenamefont
  {Schmickler}}]{KornyshevElectronPlane}\BibitemOpen
  \bibfield  {author} {\bibinfo {author} {\bibfnamefont {A.~A.}\ \bibnamefont
  {Kornyshev}}, \bibinfo {author} {\bibfnamefont {N.~B.}\ \bibnamefont
  {Luque}},\ and\ \bibinfo {author} {\bibfnamefont {W.}~\bibnamefont
  {Schmickler}},\ }\href {https://doi.org/10.1007/s10008-013-2316-8} {\bibfield
   {journal} {\bibinfo  {journal} {J. Solid State Electrochem.}\ }\textbf
  {\bibinfo {volume} {18}},\ \bibinfo {pages} {1345} (\bibinfo {year}
  {2014})}\BibitemShut {NoStop}\bibitem [{\citenamefont {Ruzanov}\ \emph {et~al.}(2018)\citenamefont
  {Ruzanov}, \citenamefont {Lembinen}, \citenamefont {Jakovits}, \citenamefont
  {Srirama}, \citenamefont {Voroshylova}, \citenamefont {Cordeiro},
  \citenamefont {Pereira}, \citenamefont {Rossmeisl},\ and\ \citenamefont
  {Ivaništšev}}]{RuzanovIonicLiquidCap}\BibitemOpen
  \bibfield  {author} {\bibinfo {author} {\bibfnamefont {A.}~\bibnamefont
  {Ruzanov}}, \bibinfo {author} {\bibfnamefont {M.}~\bibnamefont {Lembinen}},
  \bibinfo {author} {\bibfnamefont {P.}~\bibnamefont {Jakovits}}, \bibinfo
  {author} {\bibfnamefont {S.~N.}\ \bibnamefont {Srirama}}, \bibinfo {author}
  {\bibfnamefont {I.~V.}\ \bibnamefont {Voroshylova}}, \bibinfo {author}
  {\bibfnamefont {M.~N. D.~S.}\ \bibnamefont {Cordeiro}}, \bibinfo {author}
  {\bibfnamefont {C.~M.}\ \bibnamefont {Pereira}}, \bibinfo {author}
  {\bibfnamefont {J.}~\bibnamefont {Rossmeisl}},\ and\ \bibinfo {author}
  {\bibfnamefont {V.~B.}\ \bibnamefont {Ivaništšev}},\ }\href
  {https://doi.org/10.1039/C7CP07939G} {\bibfield  {journal} {\bibinfo
  {journal} {Phys. Chem. Chem. Phys.}\ }\textbf {\bibinfo {volume} {20}},\
  \bibinfo {pages} {10275} (\bibinfo {year} {2018})}\BibitemShut {NoStop}\bibitem [{\citenamefont {Paek}\ \emph {et~al.}(2013)\citenamefont {Paek},
  \citenamefont {Pak},\ and\ \citenamefont {Hwang}}]{PaekCombineDFT-MD}\BibitemOpen
  \bibfield  {author} {\bibinfo {author} {\bibfnamefont {E.}~\bibnamefont
  {Paek}}, \bibinfo {author} {\bibfnamefont {A.~J.}\ \bibnamefont {Pak}},\ and\
  \bibinfo {author} {\bibfnamefont {G.~S.}\ \bibnamefont {Hwang}},\ }\href
  {https://doi.org/10.1149/2.019301jes} {\bibfield  {journal} {\bibinfo
  {journal} {J. Electrochem. Soc.}\ }\textbf {\bibinfo {volume} {160}},\
  \bibinfo {pages} {A1} (\bibinfo {year} {2013})}\BibitemShut {NoStop}\bibitem [{\citenamefont {Willard}\ \emph {et~al.}(2009)\citenamefont
  {Willard}, \citenamefont {Reed}, \citenamefont {Madden},\ and\ \citenamefont
  {Chandler}}]{adamPot}\BibitemOpen
  \bibfield  {author} {\bibinfo {author} {\bibfnamefont {A.~P.}\ \bibnamefont
  {Willard}}, \bibinfo {author} {\bibfnamefont {S.~K.}\ \bibnamefont {Reed}},
  \bibinfo {author} {\bibfnamefont {P.~A.}\ \bibnamefont {Madden}},\ and\
  \bibinfo {author} {\bibfnamefont {D.}~\bibnamefont {Chandler}},\ }\href
  {https://doi.org/https://doi.org/10.1039/B805544K} {\bibfield  {journal}
  {\bibinfo  {journal} {Faraday Discuss.}\ }\textbf {\bibinfo {volume} {141}},\
  \bibinfo {pages} {423} (\bibinfo {year} {2009})}\BibitemShut {NoStop}\bibitem [{\citenamefont {Hansen}\ and\ \citenamefont
  {Rossmeisl}(2016)}]{Rossmeisl}\BibitemOpen
  \bibfield  {author} {\bibinfo {author} {\bibfnamefont {M.~H.}\ \bibnamefont
  {Hansen}}\ and\ \bibinfo {author} {\bibfnamefont {J.}~\bibnamefont
  {Rossmeisl}},\ }\href
  {https://doi.org/https://doi.org/10.1021/acs.jpcc.6b09019} {\bibfield
  {journal} {\bibinfo  {journal} {J. Phys. Chem. C}\ }\textbf {\bibinfo
  {volume} {120}},\ \bibinfo {pages} {29135–29143} (\bibinfo {year}
  {2016})}\BibitemShut {NoStop}\bibitem [{\citenamefont {Le}\ \emph {et~al.}(2017)\citenamefont {Le},
  \citenamefont {Iannuzzi}, \citenamefont {Cuesta},\ and\ \citenamefont
  {Cheng}}]{Cheng}\BibitemOpen
  \bibfield  {author} {\bibinfo {author} {\bibfnamefont {J.}~\bibnamefont
  {Le}}, \bibinfo {author} {\bibfnamefont {M.}~\bibnamefont {Iannuzzi}},
  \bibinfo {author} {\bibfnamefont {A.}~\bibnamefont {Cuesta}},\ and\ \bibinfo
  {author} {\bibfnamefont {J.}~\bibnamefont {Cheng}},\ }\href
  {https://doi.org/10.1103/PhysRevLett.119.016801} {\bibfield  {journal}
  {\bibinfo  {journal} {Phys. Rev. Lett.}\ }\textbf {\bibinfo {volume} {119}},\
  \bibinfo {pages} {016801} (\bibinfo {year} {2017})}\BibitemShut {NoStop}\bibitem [{\citenamefont {Valette}(1982)}]{Valette}\BibitemOpen
  \bibfield  {author} {\bibinfo {author} {\bibnamefont {Valette}},\ }\href
  {https://doi.org/10.1016/0022-0728(82)87126-X} {\bibfield  {journal}
  {\bibinfo  {journal} {J. Electroanal. Chem. Interf. Electrochem.}\ }\textbf
  {\bibinfo {volume} {138}},\ \bibinfo {pages} {37} (\bibinfo {year}
  {1982})}\BibitemShut {NoStop}\bibitem [{\citenamefont {Ojha}\ \emph {et~al.}(2020)\citenamefont {Ojha},
  \citenamefont {Arulmozhi}, \citenamefont {Aranzales},\ and\ \citenamefont
  {Koper}}]{KoperPt}\BibitemOpen
  \bibfield  {author} {\bibinfo {author} {\bibfnamefont {K.}~\bibnamefont
  {Ojha}}, \bibinfo {author} {\bibfnamefont {N.}~\bibnamefont {Arulmozhi}},
  \bibinfo {author} {\bibfnamefont {D.}~\bibnamefont {Aranzales}},\ and\
  \bibinfo {author} {\bibfnamefont {M.~T.~M.}\ \bibnamefont {Koper}},\ }\href
  {https://doi.org/https://doi.org/10.1002/ange.201911929} {\bibfield
  {journal} {\bibinfo  {journal} {Angew. Chem.}\ }\textbf {\bibinfo {volume}
  {59}},\ \bibinfo {pages} {711} (\bibinfo {year} {2020})}\BibitemShut
  {NoStop}\bibitem [{\citenamefont {Georgi}\ \emph {et~al.}(2010)\citenamefont {Georgi},
  \citenamefont {A.A.Kornyshev},\ and\ \citenamefont {M.V.Fedorov}}]{Estrict}\BibitemOpen
  \bibfield  {author} {\bibinfo {author} {\bibfnamefont {N.}~\bibnamefont
  {Georgi}}, \bibinfo {author} {\bibnamefont {A.A.Kornyshev}},\ and\ \bibinfo
  {author} {\bibnamefont {M.V.Fedorov}},\ }\href
  {https://doi.org/https://doi.org/10.1016/j.jelechem.2010.07.004} {\bibfield
  {journal} {\bibinfo  {journal} {Journal of Electroanalytical Chemistry}\
  }\textbf {\bibinfo {volume} {649}},\ \bibinfo {pages} {261} (\bibinfo {year}
  {2010})}\BibitemShut {NoStop}\bibitem [{\citenamefont {Vanzo}\ \emph {et~al.}(2014)\citenamefont {Vanzo},
  \citenamefont {Bratko},\ and\ \citenamefont {Luzar}}]{Electrostriction}\BibitemOpen
  \bibfield  {author} {\bibinfo {author} {\bibfnamefont {D.}~\bibnamefont
  {Vanzo}}, \bibinfo {author} {\bibfnamefont {D.}~\bibnamefont {Bratko}},\ and\
  \bibinfo {author} {\bibfnamefont {A.}~\bibnamefont {Luzar}},\ }\href
  {https://doi.org/https://doi.org/10.1063/1.4865126} {\bibfield  {journal}
  {\bibinfo  {journal} {J. Chem. Phys.}\ }\textbf {\bibinfo {volume} {140}},\
  \bibinfo {pages} {074710} (\bibinfo {year} {2014})}\BibitemShut {NoStop}\bibitem [{\citenamefont {Plimpton}(1995)}]{LAMMPS}\BibitemOpen
  \bibfield  {author} {\bibinfo {author} {\bibfnamefont {S.}~\bibnamefont
  {Plimpton}},\ }\href {https://lammps.sandia.gov} {\bibfield  {journal}
  {\bibinfo  {journal} {Journal of Computational Physics}\ }\textbf {\bibinfo
  {volume} {117}},\ \bibinfo {pages} {1} (\bibinfo {year} {1995})}\BibitemShut
  {NoStop}\bibitem [{\citenamefont {Berendsen}\ \emph {et~al.}(1987)\citenamefont
  {Berendsen}, \citenamefont {Grigera},\ and\ \citenamefont
  {Straatsma}}]{SPCE}\BibitemOpen
  \bibfield  {author} {\bibinfo {author} {\bibfnamefont {H.~J.}\ \bibnamefont
  {Berendsen}}, \bibinfo {author} {\bibfnamefont {J.~R.}\ \bibnamefont
  {Grigera}},\ and\ \bibinfo {author} {\bibfnamefont {T.~P.}\ \bibnamefont
  {Straatsma}},\ }\href {https://doi.org/10.1021/j100308a038} {\bibfield
  {journal} {\bibinfo  {journal} {J. Phys. Chem.}\ }\textbf {\bibinfo {volume}
  {91}},\ \bibinfo {pages} {6269} (\bibinfo {year} {1987})}\BibitemShut
  {NoStop}\bibitem [{\citenamefont {Ryckaert}\ \emph {et~al.}(1977)\citenamefont
  {Ryckaert}, \citenamefont {Ciccotti},\ and\ \citenamefont
  {Berendsen}}]{SHAKE1}\BibitemOpen
  \bibfield  {author} {\bibinfo {author} {\bibfnamefont {J.-P.}\ \bibnamefont
  {Ryckaert}}, \bibinfo {author} {\bibfnamefont {G.}~\bibnamefont {Ciccotti}},\
  and\ \bibinfo {author} {\bibfnamefont {H.~J.~C.}\ \bibnamefont {Berendsen}},\
  }\href {https://doi.org/https://doi.org/10.1016/0021-9991(77)90098-5}
  {\bibfield  {journal} {\bibinfo  {journal} {Journal of Computational
  Physics}\ }\textbf {\bibinfo {volume} {23}},\ \bibinfo {pages} {327}
  (\bibinfo {year} {1977})}\BibitemShut {NoStop}\bibitem [{\citenamefont {Andersen}(1983)}]{SHAKE2}\BibitemOpen
  \bibfield  {author} {\bibinfo {author} {\bibfnamefont {H.~C.}\ \bibnamefont
  {Andersen}},\ }\href
  {https://doi.org/https://doi.org/10.1016/0021-9991(83)90014-1} {\bibfield
  {journal} {\bibinfo  {journal} {Journal of Computational Physics}\ }\textbf
  {\bibinfo {volume} {52}},\ \bibinfo {pages} {24} (\bibinfo {year}
  {1983})}\BibitemShut {NoStop}\bibitem [{\citenamefont {Fyta}\ and\ \citenamefont {Netz}(2012)}]{Fyta2012}\BibitemOpen
  \bibfield  {author} {\bibinfo {author} {\bibfnamefont {M.}~\bibnamefont
  {Fyta}}\ and\ \bibinfo {author} {\bibfnamefont {R.~R.}\ \bibnamefont
  {Netz}},\ }\href {https://doi.org/10.1063/1.3693330} {\bibfield  {journal}
  {\bibinfo  {journal} {The Journal of Chemical Physics}\ }\textbf {\bibinfo
  {volume} {136}},\ \bibinfo {pages} {124103} (\bibinfo {year}
  {2012})}\BibitemShut {NoStop}\bibitem [{\citenamefont {Chen}\ and\ \citenamefont
  {Smith}(2007)}]{SPCESurfaceTension}\BibitemOpen
  \bibfield  {author} {\bibinfo {author} {\bibfnamefont {F.}~\bibnamefont
  {Chen}}\ and\ \bibinfo {author} {\bibfnamefont {P.~E.}\ \bibnamefont
  {Smith}},\ }\href {https://doi.org/https://doi.org/10.1063/1.2745718}
  {\bibfield  {journal} {\bibinfo  {journal} {J. Chem. Phys. 126}\ }\textbf
  {\bibinfo {volume} {126}},\ \bibinfo {pages} {221101} (\bibinfo {year}
  {2007})}\BibitemShut {NoStop}\bibitem [{\citenamefont {Merlet}\ \emph {et~al.}(2013)\citenamefont {Merlet},
  \citenamefont {Péan}, \citenamefont {Rotenberg}, \citenamefont {Madden},
  \citenamefont {Simon},\ and\ \citenamefont {Salanne}}]{Salanne}\BibitemOpen
  \bibfield  {author} {\bibinfo {author} {\bibfnamefont {C.}~\bibnamefont
  {Merlet}}, \bibinfo {author} {\bibfnamefont {C.}~\bibnamefont {Péan}},
  \bibinfo {author} {\bibfnamefont {B.}~\bibnamefont {Rotenberg}}, \bibinfo
  {author} {\bibfnamefont {P.~A.}\ \bibnamefont {Madden}}, \bibinfo {author}
  {\bibfnamefont {P.}~\bibnamefont {Simon}},\ and\ \bibinfo {author}
  {\bibfnamefont {M.}~\bibnamefont {Salanne}},\ }\href
  {https://doi.org/https://doi.org/10.1021/jz3019226} {\bibfield  {journal}
  {\bibinfo  {journal} {J. Phys. Chem. Lett.}\ }\textbf {\bibinfo {volume}
  {4}},\ \bibinfo {pages} {264} (\bibinfo {year} {2013})}\BibitemShut {NoStop}\bibitem [{\citenamefont {Reed}\ \emph {et~al.}(2008)\citenamefont {Reed},
  \citenamefont {Madden},\ and\ \citenamefont {Papadopoulos}}]{ReedMadelung}\BibitemOpen
  \bibfield  {author} {\bibinfo {author} {\bibfnamefont {S.~K.}\ \bibnamefont
  {Reed}}, \bibinfo {author} {\bibfnamefont {P.~A.}\ \bibnamefont {Madden}},\
  and\ \bibinfo {author} {\bibfnamefont {A.}~\bibnamefont {Papadopoulos}},\
  }\href {https://doi.org/10.1063/1.2844801} {\bibfield  {journal} {\bibinfo
  {journal} {J. Chem. Phys.}\ }\textbf {\bibinfo {volume} {128}},\ \bibinfo
  {pages} {124701} (\bibinfo {year} {2008})}\BibitemShut {NoStop}\bibitem [{\citenamefont {Kathmann}\ \emph {et~al.}(2011)\citenamefont
  {Kathmann}, \citenamefont {Kuo}, \citenamefont {Mundy},\ and\ \citenamefont
  {Schenter}}]{WaterSurfPot}\BibitemOpen
  \bibfield  {author} {\bibinfo {author} {\bibfnamefont {S.~M.}\ \bibnamefont
  {Kathmann}}, \bibinfo {author} {\bibfnamefont {I.-F.~W.}\ \bibnamefont
  {Kuo}}, \bibinfo {author} {\bibfnamefont {C.~J.}\ \bibnamefont {Mundy}},\
  and\ \bibinfo {author} {\bibfnamefont {G.~K.}\ \bibnamefont {Schenter}},\
  }\href@noop {} {\bibfield  {journal} {\bibinfo  {journal} {J. Phys. Chem. B}\
  }\textbf {\bibinfo {volume} {115}},\ \bibinfo {pages} {4369–4377} (\bibinfo
  {year} {2011})}\BibitemShut {NoStop}\bibitem [{\citenamefont {Sundararaman}\ \emph {et~al.}(2017)\citenamefont
  {Sundararaman}, \citenamefont {Letchworth-Weaver}, \citenamefont {Schwarz},
  \citenamefont {Gunceler}, \citenamefont {Ozhabes},\ and\ \citenamefont
  {Arias}}]{JDFTx}\BibitemOpen
  \bibfield  {author} {\bibinfo {author} {\bibfnamefont {R.}~\bibnamefont
  {Sundararaman}}, \bibinfo {author} {\bibfnamefont {K.}~\bibnamefont
  {Letchworth-Weaver}}, \bibinfo {author} {\bibfnamefont {K.}~\bibnamefont
  {Schwarz}}, \bibinfo {author} {\bibfnamefont {D.}~\bibnamefont {Gunceler}},
  \bibinfo {author} {\bibfnamefont {Y.}~\bibnamefont {Ozhabes}},\ and\ \bibinfo
  {author} {\bibfnamefont {T.~A.}\ \bibnamefont {Arias}},\ }\href@noop {}
  {\bibfield  {journal} {\bibinfo  {journal} {SoftwareX}\ }\textbf {\bibinfo
  {volume} {6}},\ \bibinfo {pages} {278} (\bibinfo {year} {2017})}\BibitemShut
  {NoStop}\bibitem [{\citenamefont {Perdew}\ \emph {et~al.}(1996)\citenamefont {Perdew},
  \citenamefont {Burke},\ and\ \citenamefont {Ernzerhof}}]{PBE}\BibitemOpen
  \bibfield  {author} {\bibinfo {author} {\bibfnamefont {J.~P.}\ \bibnamefont
  {Perdew}}, \bibinfo {author} {\bibfnamefont {K.}~\bibnamefont {Burke}},\ and\
  \bibinfo {author} {\bibfnamefont {M.}~\bibnamefont {Ernzerhof}},\ }\href@noop
  {} {\bibfield  {journal} {\bibinfo  {journal} {Phys. Rev. Lett.}\ }\textbf
  {\bibinfo {volume} {77}},\ \bibinfo {pages} {3865} (\bibinfo {year}
  {1996})}\BibitemShut {NoStop}\bibitem [{\citenamefont {Garrity}\ \emph {et~al.}(2014)\citenamefont
  {Garrity}, \citenamefont {Bennett}, \citenamefont {Rabe},\ and\ \citenamefont
  {Vanderbilt}}]{GBRV}\BibitemOpen
  \bibfield  {author} {\bibinfo {author} {\bibfnamefont {K.~F.}\ \bibnamefont
  {Garrity}}, \bibinfo {author} {\bibfnamefont {J.~W.}\ \bibnamefont
  {Bennett}}, \bibinfo {author} {\bibfnamefont {K.~M.}\ \bibnamefont {Rabe}},\
  and\ \bibinfo {author} {\bibfnamefont {D.}~\bibnamefont {Vanderbilt}},\
  }\href {https://doi.org/https://doi.org/10.1016/j.commatsci.2013.08.053}
  {\bibfield  {journal} {\bibinfo  {journal} {Comput. Mater. Sci.}\ }\textbf
  {\bibinfo {volume} {81}},\ \bibinfo {pages} {446 } (\bibinfo {year}
  {2014})}\BibitemShut {NoStop}\bibitem [{\citenamefont {Sundararaman}\ and\ \citenamefont
  {Arias}(2013)}]{TruncatedEXX}\BibitemOpen
  \bibfield  {author} {\bibinfo {author} {\bibfnamefont {R.}~\bibnamefont
  {Sundararaman}}\ and\ \bibinfo {author} {\bibfnamefont {T.}~\bibnamefont
  {Arias}},\ }\href@noop {} {\bibfield  {journal} {\bibinfo  {journal} {Phys.
  Rev. B}\ }\textbf {\bibinfo {volume} {87}},\ \bibinfo {pages} {165122}
  (\bibinfo {year} {2013})}\BibitemShut {NoStop}\bibitem [{\citenamefont {Le}\ and\ \citenamefont {Cheng}(2020)}]{ChengReview}\BibitemOpen
  \bibfield  {author} {\bibinfo {author} {\bibfnamefont {J.-B.}\ \bibnamefont
  {Le}}\ and\ \bibinfo {author} {\bibfnamefont {J.}~\bibnamefont {Cheng}},\
  }\href@noop {} {\bibfield  {journal} {\bibinfo  {journal} {Current Opinion in
  Electrochemistry}\ }\textbf {\bibinfo {volume} {19}},\ \bibinfo {pages} {129}
  (\bibinfo {year} {2020})}\BibitemShut {NoStop}\bibitem [{\citenamefont {Sundararaman}\ \emph {et~al.}(2015)\citenamefont
  {Sundararaman}, \citenamefont {Schwarz}, \citenamefont {Letchworth-Weaver},\
  and\ \citenamefont {Arias}}]{SaLSA}\BibitemOpen
  \bibfield  {author} {\bibinfo {author} {\bibfnamefont {R.}~\bibnamefont
  {Sundararaman}}, \bibinfo {author} {\bibfnamefont {K.}~\bibnamefont
  {Schwarz}}, \bibinfo {author} {\bibfnamefont {K.}~\bibnamefont
  {Letchworth-Weaver}},\ and\ \bibinfo {author} {\bibfnamefont {T.~A.}\
  \bibnamefont {Arias}},\ }\href {https://doi.org/10.1063/1.4906828} {\bibfield
   {journal} {\bibinfo  {journal} {J. Chem. Phys.}\ }\textbf {\bibinfo {volume}
  {142}},\ \bibinfo {pages} {054102} (\bibinfo {year} {2015})}\BibitemShut
  {NoStop}\bibitem [{\citenamefont {Sendner}\ \emph {et~al.}(2009)\citenamefont
  {Sendner}, \citenamefont {Horinek}, \citenamefont {Bocquet},\ and\
  \citenamefont {Netz}}]{NetzInterfacial}\BibitemOpen
  \bibfield  {author} {\bibinfo {author} {\bibfnamefont {C.}~\bibnamefont
  {Sendner}}, \bibinfo {author} {\bibfnamefont {D.}~\bibnamefont {Horinek}},
  \bibinfo {author} {\bibfnamefont {L.}~\bibnamefont {Bocquet}},\ and\ \bibinfo
  {author} {\bibfnamefont {R.~R.}\ \bibnamefont {Netz}},\ }\href@noop {}
  {\bibfield  {journal} {\bibinfo  {journal} {Langmuir}\ }\textbf {\bibinfo
  {volume} {25}},\ \bibinfo {pages} {10768–10781} (\bibinfo {year}
  {2009})}\BibitemShut {NoStop}\bibitem [{\citenamefont {Limmer}\ \emph
  {et~al.}(2013{\natexlab{b}})\citenamefont {Limmer}, \citenamefont {Willard},
  \citenamefont {Madden},\ and\ \citenamefont {Chandler}}]{LimmerHydrophobic}\BibitemOpen
  \bibfield  {author} {\bibinfo {author} {\bibfnamefont {D.~T.}\ \bibnamefont
  {Limmer}}, \bibinfo {author} {\bibfnamefont {A.~P.}\ \bibnamefont {Willard}},
  \bibinfo {author} {\bibfnamefont {P.}~\bibnamefont {Madden}},\ and\ \bibinfo
  {author} {\bibfnamefont {D.}~\bibnamefont {Chandler}},\ }\href
  {https://doi.org/https://doi.org/10.1073/pnas.1301596110} {\bibfield
  {journal} {\bibinfo  {journal} {PNAS}\ }\textbf {\bibinfo {volume} {110}},\
  \bibinfo {pages} {4200} (\bibinfo {year} {2013}{\natexlab{b}})}\BibitemShut
  {NoStop}\bibitem [{\citenamefont {Hockney}\ and\ \citenamefont {W.}(1988)}]{pppm}\BibitemOpen
  \bibfield  {author} {\bibinfo {author} {\bibfnamefont {R.~W.}\ \bibnamefont
  {Hockney}}\ and\ \bibinfo {author} {\bibfnamefont {E.~J.}\ \bibnamefont
  {W.}},\ }\href {https://doi.org/10.1201/9780367806934} {\emph {\bibinfo
  {title} {Computer Simulation using Particles}}}\ (\bibinfo  {publisher}
  {Taylor \& Francis},\ \bibinfo {year} {1988})\BibitemShut {NoStop}\bibitem [{\citenamefont {Grimme}(2006)}]{DFT-D2}\BibitemOpen
  \bibfield  {author} {\bibinfo {author} {\bibfnamefont {S.}~\bibnamefont
  {Grimme}},\ }\href@noop {} {\bibfield  {journal} {\bibinfo  {journal} {J.
  Comput. Chem.}\ }\textbf {\bibinfo {volume} {27}},\ \bibinfo {pages} {1787}
  (\bibinfo {year} {2006})}\BibitemShut {NoStop}\bibitem [{\citenamefont {Virtanen}\ \emph {et~al.}(2020)\citenamefont
  {Virtanen}, \citenamefont {Gommers} \emph {et~al.}}]{SciPy}\BibitemOpen
  \bibfield  {author} {\bibinfo {author} {\bibfnamefont {P.}~\bibnamefont
  {Virtanen}}, \bibinfo {author} {\bibfnamefont {R.}~\bibnamefont {Gommers}},
  \emph {et~al.},\ }\href {https://doi.org/10.1038/s41592-019-0686-2}
  {\bibfield  {journal} {\bibinfo  {journal} {Nature Methods}\ }\textbf
  {\bibinfo {volume} {17}},\ \bibinfo {pages} {261} (\bibinfo {year}
  {2020})}\BibitemShut {NoStop}\end{thebibliography}
 \end{document}